\documentclass[epj]{svjour}
\usepackage{graphics}
\usepackage[dvipdfm]{graphicx}
\def\simge{\mathrel{%
       \rlap{\raise 0.511ex \hbox{$>$}}{\lower 0.511ex \hbox{$\sim$}}}}
\def\simle{\mathrel{
       \rlap{\raise 0.511ex \hbox{$<$}}{\lower 0.511ex \hbox{$\sim$}}}}
\begin{document}
\title{Phase structure of hot dense QCD by a histogram method
\thanks{This report is based on the collaboration with the members of 
the WHOT-QCD Collaboration, N. Yamada and H. Yoneyama 
\cite{ejiri07,ejiri08,whot10,ejiri09,whot11,whot12,whotlat12,yama12}.}}
\author{Shinji Ejiri
}                     
%
%
\institute{
Graduate School of Science and Technology, Niigata University, Niigata 950-2181, Japan
}
\date{Received: date / Revised version: date}
%
\abstract{
We study the phase structure of QCD at high temperature and density by lattice QCD simulations adopting a histogram method. 
The quark mass dependence and the chemical potential dependence of the nature of phase transition are investigated focusing on the probability distribution function (histogram).
The shape of the distribution function changes with the quark mass and chemical potential. 
Through the shape of the distribution, the critical surface which separates the first order transition and crossover regions in the heavy quark region is determined for the (2+1)-flavor case. 
Moreover, we determined the critical point at finite density for two-flavor QCD with an intermediate quark mass, using a Gaussian approximation of the complex phase distribution of the quark determinant. 
The chemical potential dependence of the critical quark mass is also evaluated in the situation where two light quarks and many massive quarks exist. 
We find that the first order transition region becomes wider with the chemical potential in the many-flavor QCD.
} 
\maketitle

\section{Introduction}
\label{intro}

\begin{figure}[t]
\centerline{
\includegraphics[width=42mm]{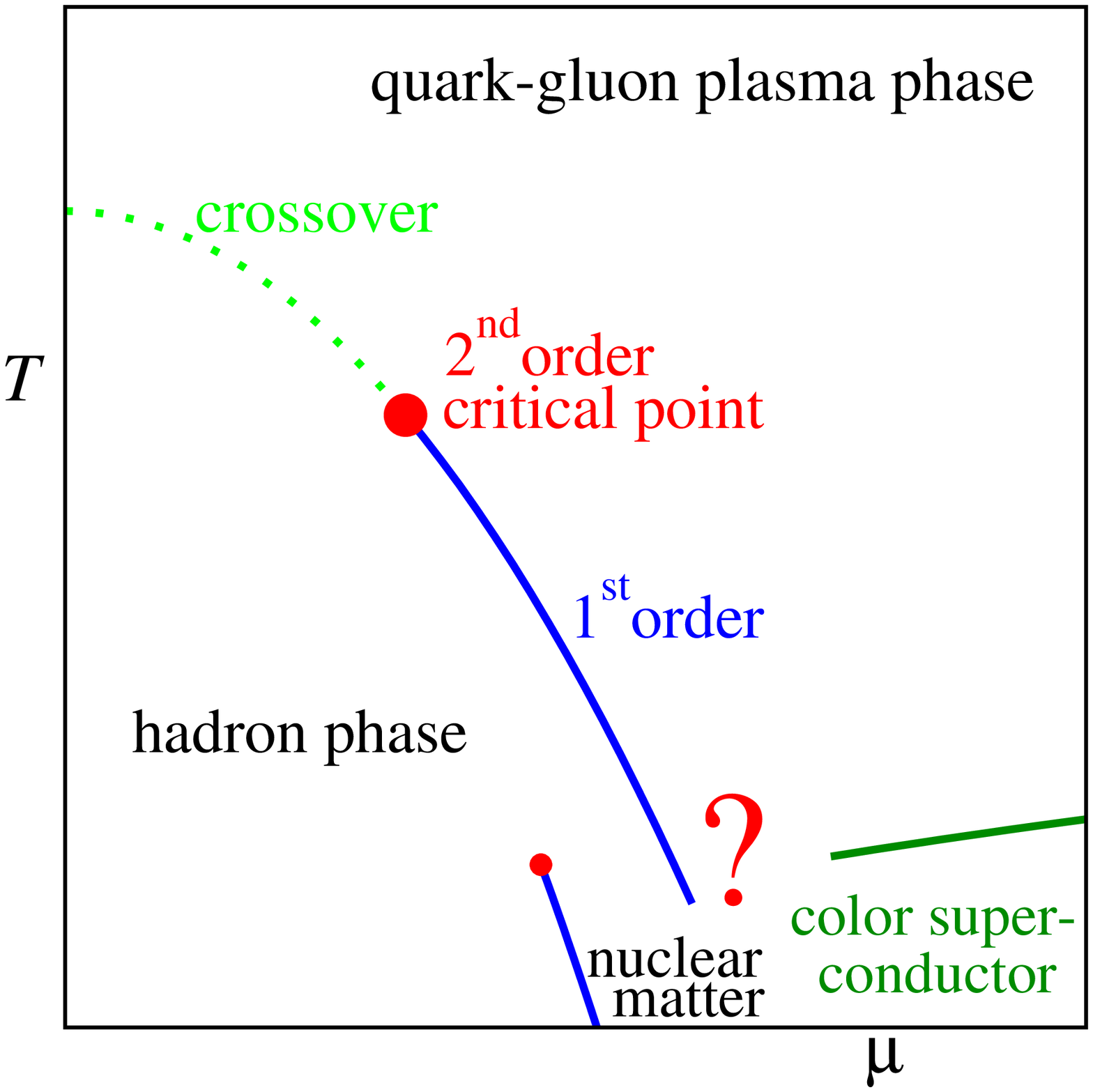}
\hskip 1mm
\includegraphics[width=42mm]{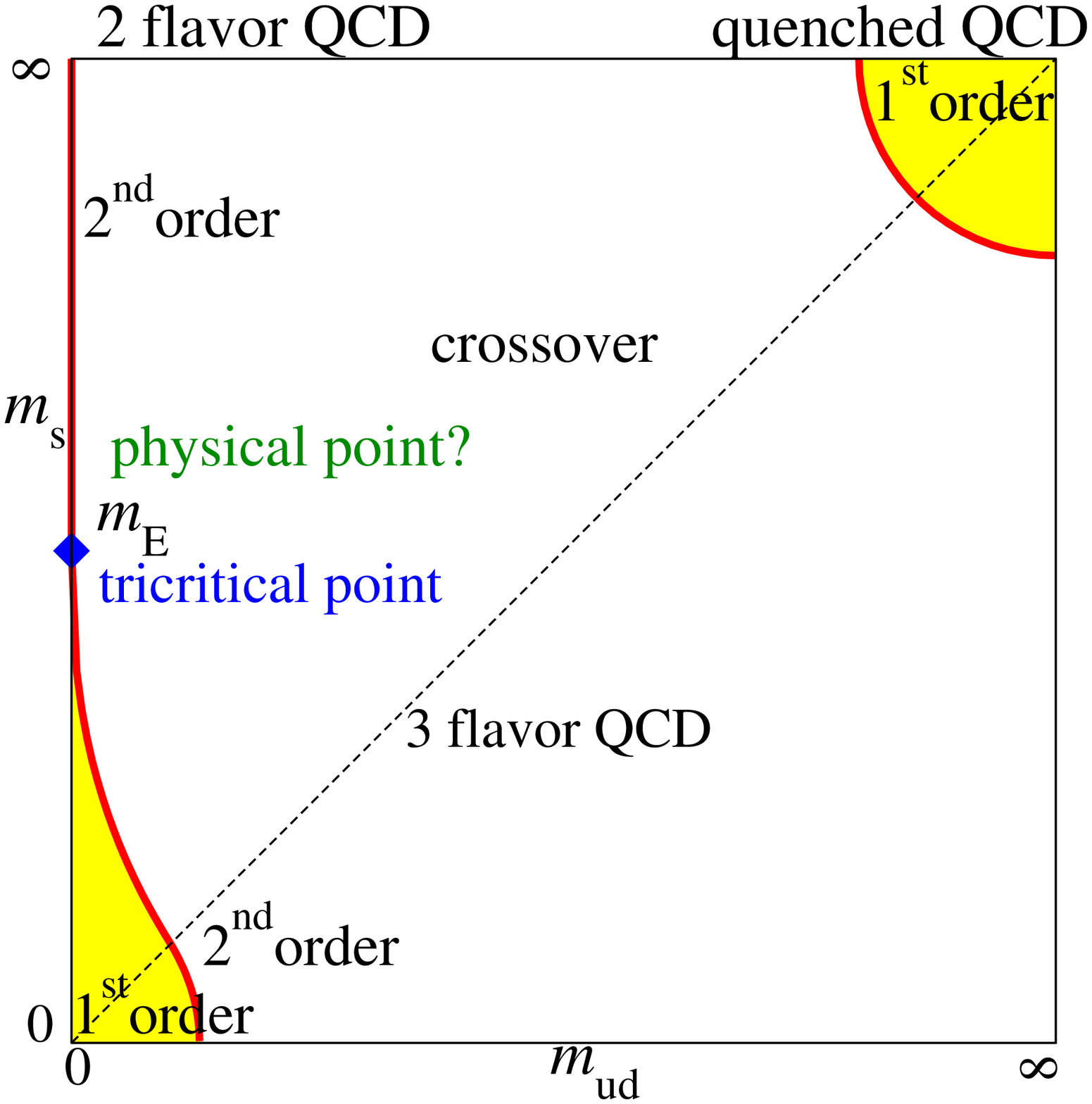}
}
\caption{ Phase diagram in the $(T, \mu)$ plane (left) 
and quark mass dependence of the order of phase transitions (right). 
}
\label{fig:massdep}
\end{figure}

The study of the QCD phase structure at high temperature and density is currently one of the most active research fields in particle physics. 
The QCD phase transition at finite temperature $(T)$ and quark chemical potential $(\mu)$ is expected to be a rapid crossover in the low density regime \cite{Wupper2006,RBCBi09}, and changes into a first order phase transition beyond a critical value of the quark number density. 
We discuss the critical point terminating the first order phase transition line in the $(T, \mu)$ phase diagram sketched in the left panel of Fig.~\ref{fig:massdep}.
The critical point is one of the most interesting features that may be discovered in heavy-ion collision experiments. 
In this paper, we summarize a series of studies about the critical point in the QCD phase diagram using a histogram method. 

To find the critical point, it is helpful to investigate the phase transition changing the quark masses as well as the chemical potential \cite{crtpt02,crtpt03,crtpt04,dFP03,dFP07}.
The order of the phase transition depends on the quark masses for $(2+1)$-flavor QCD including dynamical up, down and strange quarks. 
Although the study at finite density is difficult, by changing the quark mass, the critical point at finite density can be shifted to the low density regime, where we can study it by a simulation. 

The expected nature of the phase transition at $\mu=0$ is summarized in the right panel of Fig.~\ref{fig:massdep} \cite{kanaya10}. 
The horizontal axis $m_{ud}$ is the up and down quark masses and the vertical axis $m_s$ is the strange quark mass. 
The phase transition of 3-flavor QCD in the chiral limit, $(m_{ud}, m_s)=(0,0)$, is of first order 
\cite{Pisarski83} and is also first order in the quenched limit, $(m_{ud}, m_s)=(\infty, \infty)$ \cite{FOU90,QCDPAX92}.
The boundaries separating the first order and the crossover regions in $(2+1)$-flavor QCD are second order critical lines, which is shown by the bold red line in Fig.~\ref{fig:massdep} (right).
There are two possibilities in the chiral limit of 2-flavor QCD $(m_s= \infty)$. 
The standard scenario is that the transition in the chiral limit, $(m_{ud}, m_s)=(0, \infty)$, is second order \cite{RBCBi09,Pisarski83,iwasaki97,cppacs00}, and is crossover for finite $m_{ud}$. 
Then, the nature of the transition changes at the tricritical point, $(m_{ud}, m_s)=(0,m_{\rm E})$, shown in Fig.~ \ref{fig:massdep} (right).
An alternative scenario is that the transition is first order in the chiral limit of 2-flavor QCD \cite{pisa05,pisa08} \footnote{An interesting argument about the chiral phase transition of 2-flavor QCD is given in Ref.~\cite{aoki12}.}.
In this case, we have no tricritical point. 

It is more interesting to discuss the nature of the phase transition at finite density. 
Then, the critical line in Fig.~\ref{fig:massdep} (right) becomes the critical surface separating the first order and crossover regions \cite{crtpt02,crtpt03,crtpt04,dFP03,dFP07}.
The standard expectation is that the first order region becomes wider as $\mu$ increases and thus, at the physical quark mass point, the crossover transition at low density changes to be first order at high density, as is shown in the phase diagram, Fig.~\ref{fig:massdep} (left).
However, the opposite result has been obtained in Ref.~\cite{dFP07}, which suggests that the first order region becomes narrower with $\mu$ in the vicinity of $\mu=0$. 
It is thus very important to investigate the location of the critical surface precisely in the $(m_{ud}, m_s, \mu)$ space.

This report is organized as follows.
We explain the histogram method in the next section. To study the nature of the phase transition, we define a probability distribution function and investigate the shape of the distribution function. Through the shape of the distribution, we identify the order of the phase transition. 
In Sec.~\ref{sec:heavy}, we apply the histogram method to find the location of the endpoint of the first order phase transition region in the heavy quark region \cite{whot11}, which is upper right corner of Fig.~\ref{fig:massdep} (right). The chemical potential dependence of the boundary is also discussed \cite{whot12,whotlat12}.
We then discuss the existence of the critical point at finite density in 2-flavor QCD with an intermediate quark mass in Sec.~\ref{sec:2flavor} \cite{ejiri07}. 
Toward the understanding of the light quark mass region of $(2+1)$-flavor QCD, we propose the study of $(2+N_{\rm f})$-flavor QCD with large $N_{\rm f}$, where two flavors are light and the others are heavy, in Sec.~\ref{sec:manyflavor}. 
We discuss the $N_{\rm f}$ dependence and the chemical potential dependence of the critical line which separates the first order transition and crossover regions \cite{yama12}.
Finally, in Sec.~\ref{sec:canonical}, we will comment on the canonical approach for the investigation of the probability distribution of net baryon number, which is important for the study of the baryon number fluctuation in the heavy ion collisions \cite{ejiri08}.
Section \ref{sec:summary} is the summary of this study.

\section{Histogram method}
\label{sec:hist}

One of the most primitive approaches to identify the order of the phase transition is to investigate the histogram of a typical quantity such as plaquette, Polyakov loop or chiral condensate in Monte-Caro simulations. 
If the phase transition is of first order, two different states coexist at the transition point. 
To identify the first order transition, it is useful to introduce a probability distribution function. 
By looking at the shape of the distribution function, the nature of a phase transitions can be identified.
In this section, we introduce the method to explore the phase structure of QCD \cite{ejiri07}, which may viewed as a variant of the histogram method or the density of state method \cite{FS1988,Gock88,Anag02,Ambj02,Fod03}.

We characterize a configuration by the average plaquette $P$ 
(or the gauge action) for simplification although the following discussion is 
possible using a order parameter of the phase transition such as 
the Polyakov loop or the chiral condensate too. 
The probability distribution function (histogram) of the plaquette $w(P)$ is defined by 
\begin{eqnarray}
w(P) = \int {\cal D} U \ \delta(P- \hat{P}) \prod_{f=1}^{N_{\rm f}} \det M(\kappa_f, \mu_f) 
e^{-S_g (\beta)}, 
\label{eq:pqdist}
\end{eqnarray}
where $\delta(x)$ is the delta function, 
$S_g$ is the gauge action, and $M$ is the quark matrix. 
For later discussions, we define the (generalized) average plaquette operator $\hat{P}$ as 
$ \hat{P} \equiv -S_g/(6 \beta N_{\rm site})$.
This is the average of the plaquette, $1 \times 1$ Wilson loop, over all elementary squares for the standard gauge action and is a linear combination of Wilson loops for an improved action. 
$N_{\rm site} = N_s^3 \times N_t$ is the number of sites.
$\beta=6/g_0^2$ is the lattice bare parameter. 
$\kappa$ is the hopping parameter corresponding to the inverse of the quark mass.
In the calculation of Eq.~(\ref{eq:pqdist}), we actually use an approximate delta function such as a box type function, 
$\delta(x) \approx \{ 1/\Delta \ 
({\rm for} \ \Delta/2<x \leq \Delta/2), 
0 \ {\rm (otherwise)} \}$, 
or a Gaussian function,  
$\delta(x) \approx 1/(\Delta \sqrt{\pi}) \exp[-(x/\Delta )^2]$.
For the case of the box type, we can estimate $w(P)$ by counting 
the number of configurations for each value of $P$ with 
the width of box $\Delta$. As $\Delta$ decreases, the approximation 
becomes better but the statistical error becomes 
large because the number of configurations in each block 
becomes small. 
Hence, we must adjust the size of $\Delta$ appropriately.

Here, we discuss the shape of the probability distribution function 
to study the nature of the phase transition. 
In general, the number of states increases exponentially 
as the gauge fields become random. 
On the other hand, the random configurations are exponentially 
suppressed by the weight factor $\exp(6\beta N_{\rm site} \hat{P})$, 
since the plaquette $\hat{P}$ decreases as the configuration becomes random, in general.
Therefore, the most probable $P$ is determined by 
the balance of the number of states and the weight factor,
and the value of plaquette distributes around the most 
probable value in a Monte-Carlo simulation. 

We first consider the case that there is no spatial correlation 
between the plaquette variables at each point and 
the volume is sufficiently large. 
In this case, the shape of the probability distribution 
as a function of the plaquette averaged over the space 
must be a Gaussian function. 
The central limit theorem tells us that the probability 
distribution of the average of the random numbers which have 
the same probability distribution is always Gaussian type 
when the set of random numbers is large enough, 
and the width of the distribution becomes narrower in inverse 
proportion to the square root of the number of random numbers.
We can apply this theorem to this case. Hence, 
\begin{eqnarray}
w(P)= \sqrt{\frac{6N_{\rm site}}{2\pi \chi_P}} 
\exp \left\{ -\frac{6N_{\rm site}}{2 \chi_P} 
\left( P- \langle P \rangle \right)^2 \right\}, 
\label{eq:gdis}
\end{eqnarray}
where $\langle P \rangle $ is the expectation value of $P$ and 
$\chi_P$ is the susceptibility; 
\begin{eqnarray}
\langle P \rangle &=& \int P \ w(P) \ dP, 
\label{eq:gcoe} \\
\chi_P & \equiv & 6N_{\rm site} \langle (P - \langle P \rangle)^2 \rangle
= 6N_{\rm site} \int (P - \langle P \rangle)^2 w(P) dP.
\nonumber
\end{eqnarray}
For this case, $\chi_P$ is finite in the large $N_{\rm site}$ limit, since the width of the distribution decreases as $\sim N_{\rm site}^{-1/2}$.

We expect that $w(P)$ is a Gaussian function also for more general 
interacting cases when the correlation 
length is much shorter than the size of the system. 
If we divide the space into domains which are larger than 
the correlation length and average the plaquette variables in 
these domains, the averaged plaquettes can be independent 
for each domain. When the number of domains is large, 
the distribution function as a function of the plaquette 
averaged over space must be a Gaussian function. 

However, we expect that the probability distribution function is not 
Gaussian for the following two cases. 
One is, of course, the case that the correlation length is not small 
in comparison to the size of the system because the above-mentioned 
argument cannot be applied. 
The other case is that the most probable values of plaquette 
is not unique. For this case, the whole space is separated 
into domains having different states, 
and the plaquette variables in each domain distribute around one of 
the most probable values of plaquette. 
Although, on the surface separating these domains, the most probable 
plaquette value may not be realized, the effect from the wall becomes 
smaller as the volume increases, since the effect from the wall 
increases as a function of the area of the wall. 
Consequently, the existence of the domain wall does 
not affect the probability in the infinite volume limit. 
The probability distribution function should then be flat 
in the range between these most probable values of $P$ because 
the spatial average of $P$ depends on the size of these domains but 
the probability does not change in this range. 
However, in a finite volume, the effect from the domain wall 
cannot be neglected, hence the distribution function has two peaks 
when the number of most probable values of $P$ is two. 
The two exceptions discussed here correspond to 
the cases at a second order phase transition point and 
at a first order phase transition point, respectively.

It is then convenient to introduce the effective potential defined by
\begin{eqnarray}
V_{\rm eff}(P) = -\ln w(P).
\end{eqnarray}
As discussed above, the distribution function is normally written as 
$w(P) \sim \exp \{ -(6N_{\rm site}/2 \chi_P) (P- \langle P \rangle)^2 \}$.
When one considers a Taylor expansion of $V_{\rm eff}(P)$ around the minimum 
$\langle P \rangle$, where the slope of the potential $dV_{\rm eff}/dP$ is zero, 
the effective potential is dominated by the quadratic 
term in the region near the minimum, i.e. 
the potential is a quadratic function in the vicinity of 
$\langle P \rangle$, and the second derivative (curvature) 
of $V_{\rm eff}(P)$ at $\langle P \rangle$ is related to 
the plaquette susceptibility as
\begin{eqnarray}
\frac{d^2 V_{\rm eff}}{dP^2} = \frac{6N_{\rm site}}{\chi_P}.
\label{eq:curpot}
\end{eqnarray}

A second order phase transition point is characterized by 
the slope and curvature of the effective potential. 
The slope $dV_{\rm eff}/dP$ and curvature $d^2V_{\rm eff}/dP^2$ become zero 
simultaneously at the critical point.
As given in Eq.~(\ref{eq:gcoe}), $\chi_P$ is an indicator of 
fluctuations and diverges at a second order phase transition point
in the thermodynamic limit.
When the susceptibility $\chi_P$ becomes large in the vicinity of 
a second order phase transition point, the effect from 
the quadratic term of $V_{\rm eff}(P)$ becomes small 
in comparison to the higher order terms, and then 
the distribution function deviates from a Gaussian function. 
On the other hand, in the case of a first order phase transition point, 
more than one peak exist in the distribution function. 
This means that there are more than one point which gives $dV_{\rm eff}/dP=0$, 
and the curvature of $V_{\rm eff}(P)$ is negative 
around the center of the distribution. 
Such studies have been done to confirm that the phase transition of pure SU(3) gauge theory is of first order, and the double-peaked distribution have been obtained at the transition point \cite{FOU90,QCDPAX92}.
This method is equivalent to other methods to identify the order of phase transitions by the Binder cumulant and by the Lee-Yang zero \cite{ejiri06}.

Moreover, the probability distribution function can be used for the calculation of the expectation value.
Once we obtain the distribution function of a quantity $X$, the expectation value can be rewritten as 
\begin{eqnarray}
\langle {\cal O}[X] \rangle_{(\beta, \kappa)} 
&=& \frac{1}{\cal Z} \int {\cal D} U \ {\cal O}[\hat{X}] 
\prod_{f=1}^{N_{\rm f}} \det M(\kappa_f,\mu_f) e^{-S_g} 
\nonumber \\
&& \hspace{-12mm} = \frac{1}{\cal Z} \int {\cal O}[X] \ w(X) \ d X,  
\hspace{3mm}
{\cal Z} = \int w(X) \ d X,  
\label{eq:parti}
\end{eqnarray}
for an operator ${\cal O}[\hat{X}]$ given by the operator $\hat{X}$.
Through the above equations, the distribution function is directly related to the expectation values.
In the following sections, we will discuss the change of the distribution function when the parameters $\beta$, $\kappa$, $\mu$ are shifted, using a kind of the reweighting method.

\section{QCD Phase diagram in the heavy quark region}
\label{sec:heavy}

\subsection{Polyakov loop distribution function}

In this section, we study the boundary of the first order transition region in the case when quarks are all heavy \cite{whot11,whotlat12}. We determine the boundary as a function of the chemical potential $\mu$ by measuring histograms.
Although this boundary in the heavy quark region is irrelevant to the boundary near the physical point in Fig.~\ref{fig:massdep} (right), 
this provides us with a good testing and developing ground for the method, because 
the computational burden is much lighter.

We calculate the probability distribution function of the Polyakov loop $\Omega = \Omega_{\rm R} + i \Omega_{\rm I}$.
This is the most important observable near the transition point in the heavy quark region and is the order parameter of the deconfinement transition.
The probability distribution function of the real part $\Omega_{\rm R}$ is defined by 
\begin{eqnarray}
w(\Omega_{\rm R}; \beta, \kappa_f, \mu_f) && \nonumber \\
&& \hspace{-20mm} = \int {\cal D} U \ \delta(\Omega_{\rm R}- \hat{\Omega}_{\rm R}) \ 
e^{-S_g}\ \prod_{f=1}^{N_{\rm f}} \det M(\kappa_f, \mu_f) \nonumber \\
&& \hspace{-20mm} = w(\Omega_{\rm R}, \beta, 0, 0) 
\left\langle \prod_{f=1}^{N_{\rm f}} \frac{\det M(\kappa_f, \mu_f)}{\det M(0, 0)}
\right\rangle_{(\Omega_{\rm R} {\rm fixed}; \beta)} ,
\label{eq:dist}
\end{eqnarray}
where
$N_{\rm f}$ is the number of  flavors, and 
$\langle \cdots \rangle_{(\Omega_{\rm R} {\rm fixed}; \ \beta)} \equiv 
\langle \cdots \delta (\Omega_{\rm R}- \hat{\Omega}_{\rm R})\rangle_{\beta} / 
\langle \delta(\Omega_{\rm R}-\hat{\Omega}_{\rm R}) \rangle_{\beta}$ 
means the expectation value measured with fixing the operator $\hat{\Omega}_{\rm R}$ with 
$\Omega_{\rm R}$ at $\beta$ in quenched simulations, $\kappa_f=\mu_f=0$.
Here, we use the plaquette gauge action and the standard Wilson quark action;
\begin{eqnarray}
S_g &=& -6 N_{\rm site} \beta \, \hat{P}, \\
S_q &=& 
\sum_{f=1}^{N_{\rm f}} \sum_{x,y} \bar{\psi}_x^f \, M_{xy} (\kappa_f,\mu_f) \, \psi_y^f
\\
&& \hspace{-10mm}
M_{xy} (\kappa, \mu)= 
\delta_{x,y}
\\
&& \hspace{-10mm}
-\kappa
\sum_{\mu=1}^{3} \left\{ (1-\gamma_\mu)\,U_{x,\mu}\delta_{x+\hat{\mu},y} + 
(1+\gamma_\mu)\,U^\dagger_{x-\hat{\mu},\mu}\delta_{x-\hat{\mu},y}
\right\}
\nonumber \\
&& \hspace{-10mm} 
-\kappa \left\{ e^{\mu a} (1-\gamma_4)\,U_{x,4}\delta_{x+\hat{4},y} + 
e^{- \mu a} (1+\gamma_4)\,U^\dagger_{x-\hat{4},4}\delta_{x-\hat{4},y}
\right\} . \nonumber 
\end{eqnarray}
The expectation value in the right hand side of Eq.~(\ref{eq:dist}) is the ratio of 
$w(\Omega_{\rm R},\beta, \kappa_f, \mu_f)$ and $w(\Omega_{\rm R},\beta, 0, 0)$. 
However, the calculation of $\det M$ is usually difficult. We perform the hopping parameter expansion and compute the quark determinant in the leading order of the expansion;
\begin{eqnarray}
\frac{\det M(\kappa, \mu)}{\det M(0,0)} 
&=& \exp \left[ 
288N_{\rm site} \kappa^4 \hat{P} 
+3N_{\rm s}^32^{N_{\rm t}+2} \kappa^{N_{\rm t}} \right.
\nonumber \\
&& \hspace{-10mm} \left. \times 
\left\{\cosh \left(\frac{\mu}{T}\right) \hat\Omega_{\rm R}
+i\sinh \left( \frac{\mu}{T} \right) \hat\Omega_{\rm I}\right\} 
+ \cdots \right]
\label{eq:detM}
\end{eqnarray}
for the standard Wilson quark action, where $\det M(0,0) = 1$.
The quark determinant is simply given by the average plaquette operator $\hat{P}$ and the real and imaginary parts of the Polyakov loop operator, $\hat{\Omega} = \hat{\Omega}_{\rm R} +i \hat{\Omega}_{\rm I}$.
Because the critical $\kappa$ is very small, at least for $N_t=4$, this approximation can be justified for the determination of the critical $\kappa$.

In this calculation, it is essential to perform simulations at several simulation points and to combine these data by the multi-point reweighting method \cite{FS1989}
\footnote{ In this study, we neglect the difference of the auto-correlation time among different simulations to simplify the analysis. The treatment of the auto-correlation time is discussed in Ref.~\cite{FS1989}.} . 
Since values of most of the observables distribute in a narrow range during one Monte-Carlo simulation, it is difficult to investigate the shape of the distribution function in a wide range.
We thus combine several simulations.
The expectation value of a operator $\hat{X}$ at $\beta$ is computed by simulations at $\beta_i$ with 
the number of configuration $N_i$ for $i=1, \cdots, N_{\rm SP}$ using the following equation in the case of the plaquette action and degenerate $N_{\rm f}$-flavor,
\begin{eqnarray}
\left\langle \hat{X} \right\rangle_{(\beta, \kappa)}
&=& \frac{1}{\cal Z} \int {\cal D} U \ \hat{X} 
e^{-S_g} (\det M(\kappa, \mu))^{N_{\rm f}}
\nonumber \\
&& \hspace{-15mm} = 
\frac{ \left\langle \hat{X} \hat{G}(\hat{P}) 
[\det M(\kappa, \mu) / \det M(0, 0)]^{N_{\rm f}} \right\rangle_{\rm all}
}{ \left\langle \hat{G}(\hat{P}) [\det M(\kappa, \mu) 
/ \det M(0, 0)]^{N_{\rm f}} \right\rangle_{\rm all}},
\label{eq:evmultb}
\end{eqnarray}
where the weight factor $\hat{G}(\hat{P})$ is
\begin{eqnarray}
\hat{G}(\hat{P})=\frac{e^{6 N_{\rm site} \beta \hat{P}}}{
\sum_{i=1}^{N_{\rm sp}} N_i e^{ {6 N_{\rm site} \beta_i \hat{P}}} 
{\cal Z}^{-1} (\beta_i)} ,
\end{eqnarray}
and $\left\langle \cdots \right\rangle_{\rm all}$ means the average over 
all configurations generated at all $\beta_i$ with $\kappa=\mu=0$.
The partition functions ${\cal Z} (\beta_i)$ at $\beta_i$ are parameters in 
this method and are determined by solving a consistency condition;
${\cal Z} (\beta_i) 
\approx \sum_{\rm \{ all \ conf. \}} \hat{G}(\hat{P})$ ,
numerically for each $i=1, \cdots, N_{\rm SP}$, except for an overall normalization constant. 
$\sum_{\rm \{all \ conf. \}}$ means the sum of configurations at all $\beta_i$.
(See the appendix A in Ref.~\cite{ejiri08} for details.)
Moreover, this method enables us to change $\beta$ continuously.

\begin{figure}[t] 
\centerline{
\includegraphics[width=90mm]{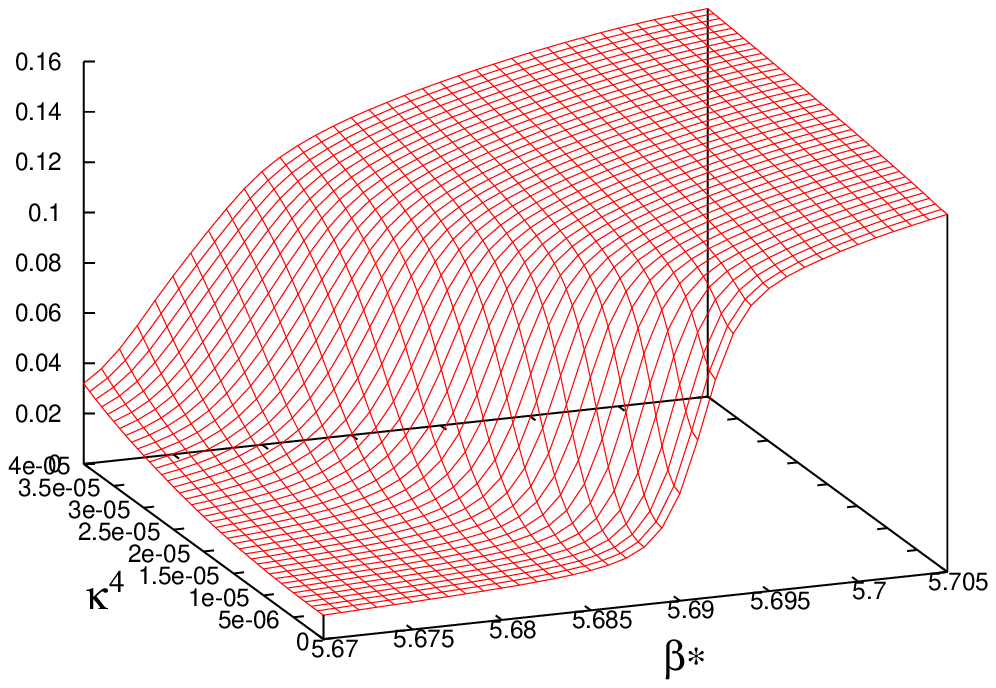} 
}
\vspace{-18mm}
\centerline{
\includegraphics[width=90mm]{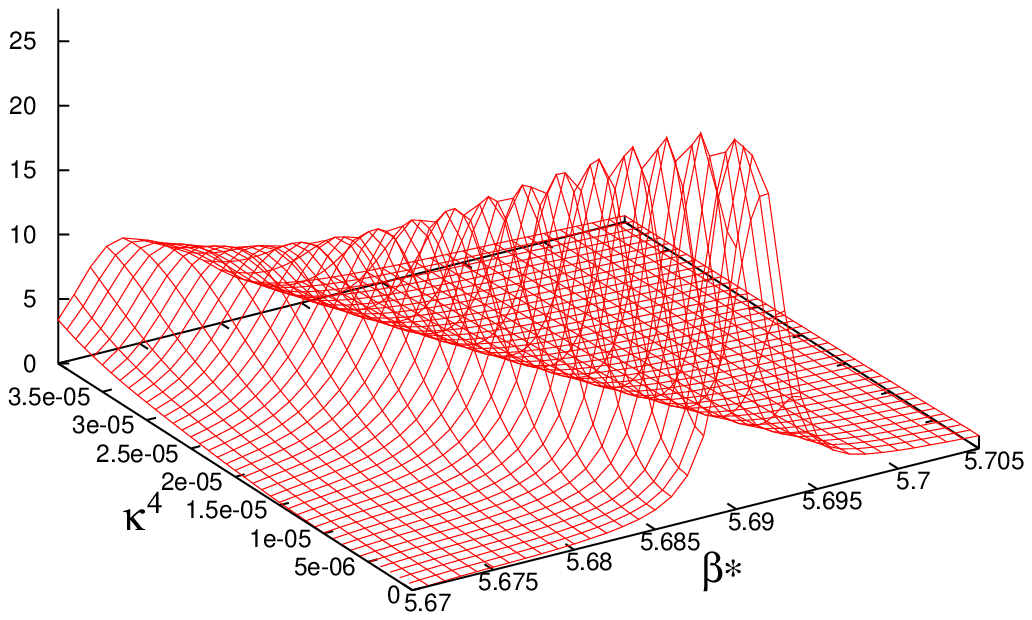} 
}
\caption{Polyakov loop (top) and its susceptibility (bottom) 
as functions of $\kappa^4$ and $\beta^* = \beta+48N_{\rm f} \kappa^4$ at $\mu=0$ for $N_{\rm f}=2$.}
\label{fig:plsus}
\hspace{0.8cm}
\end{figure}

We analyze the data obtained at 5 simulation points, $\beta =5.68$ -- $5.70$, in the quenched simulations with the plaquette gauge action on a $24^3 \times 4$ lattice \cite{whot11}. The number of configurations is 100,000 -- 670,000 for each $\beta$.
The top panel of Fig.~\ref{fig:plsus} is the result of the expectation value of the Polyakov loop and the bottom panel is its susceptibility, 
$\chi_{\Omega} =N_{s}^3 \langle (\hat{\Omega} - \langle \hat{\Omega} \rangle)^2 \rangle$,
as functions of $\kappa^{N_t}$ and the effective $\beta$ defined as $\beta^* = \beta +48N_{\rm f} \kappa^4$ with $N_t=4$, $N_{\rm f}=2$, computed at $\mu=0$ using Eq.~(\ref{eq:evmultb}).
Because the plaquette action is $S_g=-6N_{\rm site} \beta \hat{P}$, the plaquette term in Eq.~(\ref{eq:detM}) can be absorbed into the gauge action by defining $\beta^*$ and the analysis becomes simpler.
Owing to the multi-point reweighting method, the susceptibility can be calculated in a wide range of $\beta$ and $\kappa$. We define the transition point as the peak position of $\chi_{\Omega}$.

We then calculate the distribution function in heavy quark QCD for the degenerate $N_{\rm f}=2$ standard Wilson case using the multi-point reweighting method \cite{whotlat12}.
Performing the hopping parameter expansion, the distribution function can be factorized into the phase factor and the phase-quenched part; 
\begin{eqnarray}
w(\Omega_{\rm R}; \beta, \kappa, \mu) \!
&=& \! \! \! \int \! \! {\cal D} U \delta(\Omega_{\rm R} - \hat{\Omega}_{\rm R}) 
e^{6N_{\rm site} \beta \hat{P}} \! (\det M(\kappa, \mu))^{N_{\rm f}} \nonumber \\
&& \hspace{-25mm} \approx w(\Omega_{\rm R}; \beta, 0, 0) 
\left\langle  e^{288N_{\rm site} N_{\rm f} \kappa^4 \hat{P}} 
\exp \left[ 3 N_s^3 2^{N_t+2} N_{\rm f}
\kappa^{N_t} 
\right. \right. \nonumber \\  
&& \hspace{-15mm} \left. \left. \times
\left\{ \cosh \left( \frac{\mu}{T} \right) \hat{\Omega}_{\rm R}
+i\sinh \left( \frac{\mu}{T} \right) \hat{\Omega}_{\rm I} \right\} \right]
\right\rangle_{(\Omega_{\rm R} {\rm fixed}; \beta)} \nonumber \\
&& \hspace{-25mm} = w(\Omega_{\rm R}; \beta^*, 0, 0) 
\exp \left[ 3 N_s^3 2^{N_t+2}  N_{\rm f}
\kappa^{N_t} \cosh \left( \frac{\mu}{T} \right) \Omega_{\rm R} \right] 
\nonumber \\  && \hspace{-15mm} \times
\left\langle e^{i \hat{\theta}} \right\rangle_{(\Omega_{\rm R} {\rm fixed}; \beta^*)} ,
\label{eq:plhvdist}
\end{eqnarray}
where $\hat{\theta}$ is the phase of the quark determinant; 
\begin{eqnarray}
\hat{\theta} =3 N_s^3 2^{N_t+2} N_{\rm f} \kappa^{N_t} 
\sinh(\mu/T) \ \hat{\Omega_{\rm I}},
\label{eq:thetahq}
\end{eqnarray}
and the part in front of the phase average is the distribution function in the phase-quenched theory.
The plaquette term is absorbed by $S_g$ shifting $\beta$ to 
$\beta^* = \beta +48 N_{\rm f} \kappa^4$.

\begin{figure}[t] 
\centerline{
\includegraphics[width=75mm]{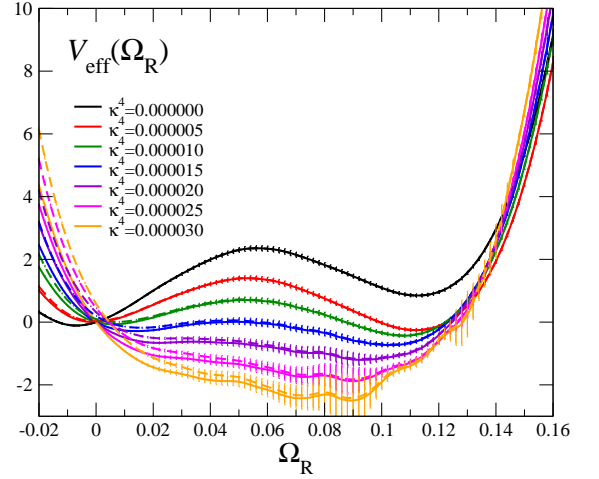}
}
\caption{The solid lines are $V_{\rm eff} (\Omega_{\rm R})$ at $\mu=0$ for each $\kappa^4$. 
$V_{\rm eff} (\Omega_{\rm R})$ at finite $\kappa^4 \cosh (\mu/T)$ is between the solid line and the dashed line \cite{whotlat12}.}
\label{fig:orhist}
\end{figure}

Expectation values with fixing the value of the Polyakov loop are computed using the delta function approximated by a Gaussian function,
$\delta(x) \approx \exp[-(x/\Delta)^2]/(\Delta \sqrt{\pi}) $, 
where $\Delta=0.005$ is adopted considering the resolution and the statistical error. 
The results of the effective potential,  
$V_{\rm eff}(\Omega_{\rm R}) = - \ln w(\Omega_{\rm R})$,
at the transition point for the case of 2-flavor QCD at $\mu=0$ are plotted by the solid lines in Fig.~\ref{fig:orhist} for several values of $\kappa^4$. 
$\beta$ is adjusted to the peak position of $\chi_{\Omega}$ at each $\kappa$.  
The value of $V_{\rm eff}(\Omega_{\rm R})$ is normalized at $\Omega_{\rm R}=0$.
This figure shows that the shape of $V_{\rm eff}(\Omega_{\rm R})$ is of double-well type at $\kappa^4=0$, indicating the first order transition,
and the shape changes gradually as increasing $\kappa$.
It becomes of single-well around $\kappa^4 \sim 0.00002$, suggesting the first order transition changes to crossover. 
The critical value of $\kappa$ has been determined by measuring the distribution function of the average plaquette in Ref.~\cite{whot11} with the same configurations.
The result is $\kappa_{\rm cp}=0.0658(3)(^{+4}_{-11})$ for $N_{\rm f}=2$.
Hence, the results of $\kappa_{\rm cp}$ from the plaquette and Polyakov loop effective potentials are consistent with each other.

The phase-quenched part of $w(\Omega_{\rm R}; \beta, \kappa, \mu)$ in Eq.~(\ref{eq:plhvdist})
can be obtained from that at $\mu=0$ simply replacing $\kappa^{N_t}$ by $\kappa^{N_t} \cosh(\mu/T)$, 
since the distribution function at $\mu=0$ is given by 
$ w(\Omega_{\rm R}, \beta^*, 0, 0) \exp [ 3 N_s^3 2^{N_t+2} N_{\rm f} \kappa^{N_t} \Omega_{\rm R} ] $.
Therefore, the critical value $\kappa_{\rm cp} (\mu)$ in the phase-quenched theory is given by 
\begin{eqnarray}
\kappa_{\rm cp}^{N_t}(0) = \kappa_{\rm cp}^{N_t}(\mu) \cosh(\mu/T).
\end{eqnarray}
The effective potential, $V_{\rm eff} (\Omega_{\rm R}) = - \ln w (\Omega_{\rm R})$, at the transition point for $\mu=0$ in Fig.~\ref{fig:orhist} is equal to the phase-quenched distribution function for each $\kappa^{N_t} \cosh(\mu/T)$.
Moreover, adopting $\kappa^{N_t} \cosh(\mu/T)$ as the basic parameter to investigate the critical point, 
the magnitude of the complex phase is limited for each $\kappa^{N_t} \cosh(\mu/T)$ 
because $\kappa^{N_t} \sinh(\mu/T)$ in $\hat{\theta}$ is always smaller than $\kappa^{N_t} \cosh(\mu/T)$.

\begin{figure}[t] 
\centerline{
\includegraphics[width=75mm]{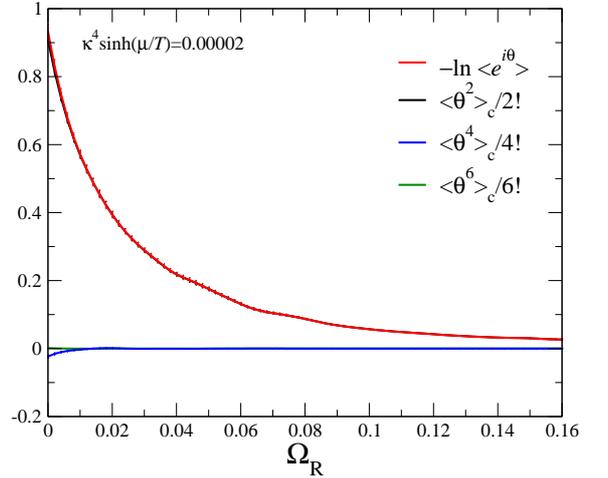} 
}
\caption{The average of the complex phase factor and the $2^{\rm nd}$, $4^{\rm th}$ and $6^{\rm th}$ order cumulants 
calculated with fixed $\Omega_{\rm R}$ at $\kappa^4 \sinh (\mu/T) = 0.00002$ \cite{whotlat12}.}
\label{fig:phase}
\end{figure}

\subsection{Complex phase of the quark determinant}
\label{sec:phase}

Next, we calculate the phase factor, 
$\langle e^{i \hat{\theta}} \rangle_{(\Omega_{\rm R} {\rm fixed}; \beta^*)}$.
If $e^{i \hat{\theta}}$ changes its sign frequently, the statistical error becomes larger than the expectation value, causing the sign problem.
To avoid the sign problem, we evaluate the phase factor by the cumulant expansion 
\cite{ejiri07,whot10}; 
\begin{eqnarray}
\left\langle e^{i \hat\theta} \right\rangle_{(\Omega_{\rm R} {\rm fixed}; \beta^*)} 
= \exp \left( \sum_{n=1}^{\infty} \frac{i^n}{n!} 
\left\langle \hat\theta^n \right\rangle_c \right).
%
\label{eq:cum}
\end{eqnarray}
where $\langle \hat\theta^n \rangle_c$ is the $n^{\rm th}$ order cumulant;
\begin{eqnarray}
\langle \hat\theta^2 \rangle_c 
&=& \langle \hat\theta^2 \rangle_{(\Omega_{\rm R})} , 
\hspace{5mm}
\langle \hat\theta^4 \rangle_c 
= \langle \hat\theta^4 \rangle_{(\Omega_{\rm R})}
-3 \langle \hat\theta^2 \rangle_{(\Omega_{\rm R})}^2, 
\nonumber \\
\langle \hat\theta^6 \rangle_c 
&=& \langle \hat\theta^6 \rangle_{(\Omega_{\rm R})}
-15 \langle \hat\theta^4 \rangle_{(\Omega_{\rm R})} 
\langle \hat\theta^2 \rangle_{(\Omega_{\rm R})} 
+30 \langle \hat\theta^2 \rangle_{(\Omega_{\rm R})}^3 
, \cdots.
\nonumber 
\end{eqnarray}
The even order terms are real and odd order terms are purely imaginary. 
The key point of this method is that $\langle \hat\theta^n \rangle_c =0$ for any odd $n$ due to the symmetry under $\hat\theta$ to $-\hat\theta$, 
i.e. $\mu$ to $-\mu$,
and thus the complex phase can be omitted from this equation.
This implies that $\langle e^{i \hat\theta} \rangle$ is guaranteed to be real and positive and the sign problem is resolved once the cumulant expansion converges.
This expansion converges in the low density region because $\hat{\theta} \sim O(\mu)$ 
and this expansion is a kind of Taylor expansion in terms of $\mu$.
Furthermore, the cumulant expansion is dominated by the lowest order of 
$\langle \hat{\theta}^2 \rangle$ if the distribution of $\hat{\theta}$ is well described by 
a Gaussian function, and a such distribution of $\hat{\theta}$ is observed in simulations with improved staggered quarks \cite{ejiri07} and with improved Wilson quarks \cite{whot10}, and has been discussed in \cite{splitt08,lomba09}.
For such a case, the phase average can be approximated by the lowest order term.

Another important point is that $\hat{\theta}$ is given by the average of the Polyakov loop.
When the correlation length is finite, the phase can be written as a summation of local 
contributions $\hat{\theta}= \sum_x \hat{\theta}_x$ with almost independent $\hat{\theta}_x$.
The phase average is then
\begin{eqnarray}
\left\langle e^{i\hat\theta} \right\rangle 
\approx \prod_x \left\langle e^{i\hat\theta_x} \right\rangle 
= \exp \left( \sum_x \sum_n \frac{i^n}{n!} 
\left\langle \hat\theta_x^n \right\rangle_c \right).
\end{eqnarray}
This suggests that all cumulants $\langle \hat\theta^n \rangle_c 
\approx \sum_x \langle \hat\theta_x^n \rangle_c$ 
increase in proportion to the volume as the volume increases.
The ratios of cumulants are thus expected to be independent of the volume 
although the sign problem becomes more serious exponentially with the volume. 
Therefore, the higher order terms in the cumulant expansion are well under control in the large volume limit.
Only for such a case, the effective potential $V_{\rm eff}$ 
can be well-defined though 
the phase-quenched effective potential $V_0$ with 
\begin{eqnarray}
V_{\rm eff} = V_0 - \ln \langle e^{i \hat{\theta}} \rangle = V_0 - \sum_n i^{n} \langle \hat{\theta}^{n} \rangle_c/n!
\end{eqnarray}
in the large volume limit, 
since $V_{\rm eff}$ and $V_0$ are both in proportion to the volume. 

We plot the results of $\langle \hat\theta^n \rangle_c /n!$ at $\kappa^{N_t} \sinh (\mu/T)=0.00002$ and $\beta^* =5.69$ in Fig.~\ref{fig:phase}.
The black, blue and green lines are the results for $n=2, 4$ and 6, respectively.
The fourth and sixth order cumulants are very small in comparison to the second order for this $\kappa$. 
The red line is $- \ln \langle e^{i \hat{\theta}} \rangle_{(\Omega_{\rm R} {\rm fixed}; \beta^*)}$, 
which is almost indistinguishable from the second order cumulant.
The contribution from the fourth and sixth orders becomes visible at small $\Omega_{\rm R}$ as $\kappa^{N_t} \sinh (\mu/T)$ increases. 
However, for the determination of the critical point, 
$\kappa_{\rm cp}^{N_t} \cosh (\mu/T) \approx 0.00002$ in the phase-quenched theory, the region at 
$\kappa^{N_t} \sinh (\mu/T) < 0.00002$ is important because $\cosh (\mu/T) > \sinh (\mu/T)$. 
This figure thus indicates that the phase average is well-approximated 
by the second order cumulant around the critical $\kappa$. 
The effective potentials including the effect from the phase factor are shown in Fig.~\ref{fig:orhist} for each $\kappa^{N_t} \cosh (\mu/T)$. 
The results at finite $\mu$ are between the solid and dashed lines.
The solid lines are the phase-quenched results, and 
the phase factors of the dashed lines are estimated by the second order cumulant when the effect from the phase factor is the largest, i.e., $\kappa^{N_t} \sinh (\mu/T)$ in Eq.~(\ref{eq:thetahq}) is equal to $\kappa^{N_t} \cosh (\mu/T)$.
We find that the contribution from the phase, 
$-\ln \langle e^{i \hat{\theta}} \rangle$, 
is quite small except at small $\Omega_{\rm R}$ and the phase factor does not affect $V_{\rm eff}$ in the region of $\Omega_{\rm R}$ relevant for the determination of the critical point. 
This means that the contribution from the complex phase to the location of the critical point is quite small on our $24^3 \times 4$ lattice.

\begin{figure}[tb]
\centerline{
\includegraphics[width=80mm]{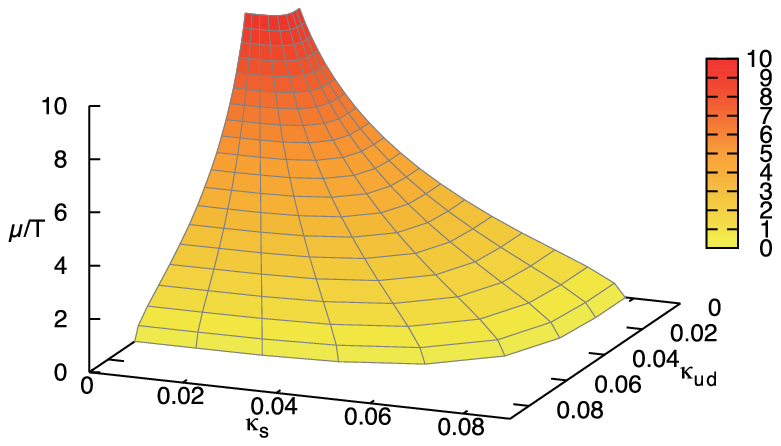}
}
\vspace{2mm}
\centerline{
\includegraphics[width=80mm]{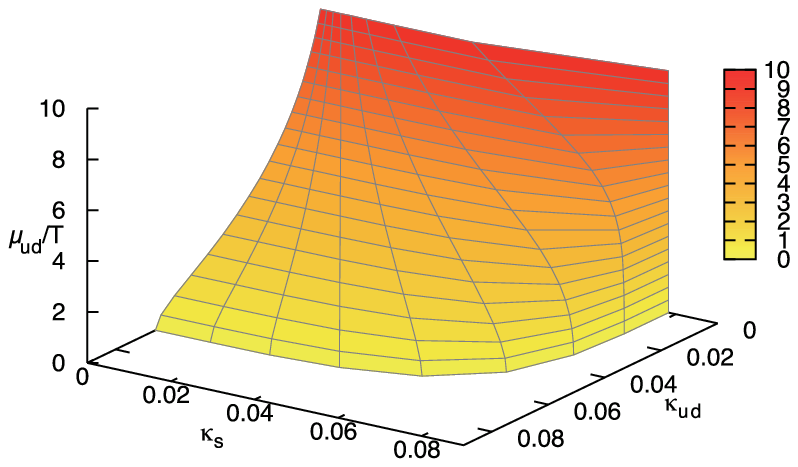}
}
\vspace{2mm}
\caption{Critical surface separating the first order transition and crossover regions in the heavy quark region \cite{whot12}.
 {\em Top:} 
The case $\mu_{u} = \mu_{d} = \mu_{s} \equiv \mu$.
 {\em Bottom:} 
The case that may be realized in heavy ion collisions: $\mu_{u} = \mu_{d} \equiv \mu_{ud}$ and $\mu_{s} = 0$.
}
\label{fig:crtsur}
\end{figure}

\subsection{(2+1)-flavor QCD in the heavy quark region}

This argument can be easily generalized to the case of non-degenerate quark masses \cite{whot12}.
For the non-degenerate $N_{\rm f}$-flavor case, the quark determinant in Eq.~(\ref{eq:plhvdist}) changes to
\begin{eqnarray}
\prod_{f=1}^{N_{\rm f}} \frac{\det M(\kappa_f, \mu)}{\det M(0,0)} 
&=& \exp \left[ 
3N_{\rm s}^3 2^{N_{\rm t}+2} \sum_{f=1}^{N_{\rm f}} \kappa_f^{N_{\rm t}} 
\right. \nonumber \\
&& \hspace{-30mm} \left. \times 
\left\{ \cosh \left( \frac{\mu_f}{T} \right) \hat{\Omega}_{\rm R}
+i \sinh \left( \frac{\mu_f}{T} \right) \hat\Omega_{\rm I} \right\} 
+ \cdots \right],
\label{eq:detMnd}
\end{eqnarray}
where the plaquette term will be absorbed into the shift of $\beta$.
For the $N_{\rm f}=2+1$ case, the difference from the $N_{\rm f}=2$ case is just to replace $2\kappa^{N_t}$ by $2 \kappa_{\rm ud}^{N_t} + \kappa_{\rm s}^{N_t}$. 
Since we have determine the critical point for $N_{\rm f}=2$ changing $\kappa$,
we find that the critical $(\kappa_{\rm ud}, \kappa_{\rm s})$ is given by 
\begin{equation}
2 \kappa_{\rm ud}^{N_t}(\mu) \cosh\left( \frac{\mu_{\rm ud}}{T} \right) 
+ \kappa_{\rm s}^{N_t}(\mu) \cosh\left( \frac{\mu_{\rm s}}{T} \right) 
= 2 [\kappa_{\rm cp}^{N_{\rm f}=2}(0) ]^{N_t} ,
\end{equation}
where we have neglected the effect of the phase factor because it does not affect the determination of the critical point, and $\kappa_{\rm cp}^{N_{\rm f}=2}(0)=0.0658(3)(^{+4}_{-11})$ for $N_t=4$ 
\cite{whot11}.
The critical lines in the $\kappa$ plane for up, down and strange are drawn 
in Fig.~\ref{fig:crtsur} 
for the cases of $\mu_{\rm ud}/T=\mu_{\rm s}/T=0$ -- $10$ (top) and 
$\mu_{\rm ud}/T=0$ -- $10$ and $\mu_{\rm s}/T=0$ (bottom). 
Because the strange chemical potential is small in 
the heavy-ion collisions, the $(2+1)$-flavor case with $\mu_{\rm h}=0$ 
corresponds to the experiments.
Note that the critical $\kappa$ decreases as $N_{\rm f}$ increases, in general.

\section{Phase diagram of two-flavor QCD}
\label{sec:2flavor}

Next, let us move on to the light quark mass region.
We discuss the distribution function of the (generalized) average plaquette $(P)$ to identify the order of the phase transition \cite{ejiri07}.
Because the number of configurations is limited for full QCD simulations in comparison with quenched QCD simulations, we use the following good properties of the plaquette distribution function to investigate the nature of the phase transition.
We define the plaquette distribution function for $N_{\rm f}$-flavor QCD with the quark masses $m_f$ and chemical potential $\mu_f$
($f=1, \cdots, N_{\rm f}$) by
\begin{eqnarray}
w(P; \beta, m_f, \mu_f) 
&=&  \int {\cal D} U {\cal D} \psi {\cal D} \bar{\psi} \
    \delta(P- \hat{P}) \ e^{- S_q - S_g} \nonumber \\
&& \hspace{-30mm} = \int {\cal D} U \ \delta(P- \hat{P}) \ 
    e^{6\beta N_{\rm site} \hat{P}}\
    \prod_{f=1}^{N_{\rm f}} (\det M(m_f, \mu_f)),
\label{eq:pdist}
\end{eqnarray}
where 
$S_g$ and $S_q$ are the gauge and quark actions, respectively. 
$M$ is the quark matrix.
$\hat P$ is the generalized plaquette operator, and
this method is applicable to the case of improved actions
replacing $\hat P$ to $\hat P=-S_g/(6N_{\rm site} \beta)$.
The partition function is ${\cal Z}= \int w(P) dP$, and 
the effective potential is then given by
\begin{eqnarray}
 V_{\rm eff}(P;\beta,m_f,\mu_f) = -\ln w(P;\beta,m_f,\mu_f).
 \label{eq:effective-potential}
\end{eqnarray}
If there is a first order phase transition point, where two different 
states coexist at the transition point, the distribution function must have two 
peak at two different values of $P$ corresponding to the hot and cold phases.
Here and hereafter, we restrict ourselves to discuss only the case when 
the quark matrix does not depend on $\beta$ explicitly
for simplicity.

We consider the ratio of the distribution function for degenerate $N_{\rm f}$-flavor QCD with fixed quark mass;
\begin{eqnarray}
R(P,\mu) & \equiv & \frac{w(P; \beta, \mu)}{w(P; \beta, 0)}
\nonumber \\
&=& \frac{\int {\cal D} U \ \delta(P- \hat{P}) (\det M(\mu))^{N_{\rm f}} e^{6\beta N_{\rm site} \hat{P}}}{
\int {\cal D} U \ \delta(P- \hat{P}) (\det M(0))^{N_{\rm f}} e^{6\beta N_{\rm site} \hat{P}}}
\nonumber \\
&=& \left\langle  
\left( \frac{\det M(\mu)}{\det M(0)} \right)^{N_{\rm f}}
\right\rangle_{(P: {\rm fixed}; \beta, 0)}.
\label{eq:rmudef}
\end{eqnarray}
Here, $\left\langle \cdots \right\rangle_{(P: {\rm fixed}; \beta, 0)}$ means 
the expectation value with fixed $P$ at $\mu=0$. 
Because we fix $P$, the factors $e^{6\beta N_{\rm site} \hat{P}}$ 
in the numerator and denominator cancel.
Therefore, this $R(P; \mu)$ is independent of $\beta$.
Although the distribution of $P$ is different for each $\beta$, 
the results obtained at different $\beta$ must be consistent.
Moreover, when we want to combine the data obtained at different $\beta$ 
using multi-point reweighting, Eq.~(\ref{eq:evmultb}), to calculate $R$ 
in a wide range of $P$, we can omit the factor $G(\hat{P})$ 
because $G(\hat{P})$ is a function of $\hat{P}$.
In this method, all simulations are performed at $\mu=0$ and the effect of 
finite $\mu$ is introduced though the operator 
$\det M(\mu) / \det M(0)$ measured on the configurations 
generated by the simulations at $\mu=0$.

\begin{figure}[t]
\centerline{
\includegraphics[width=55mm]{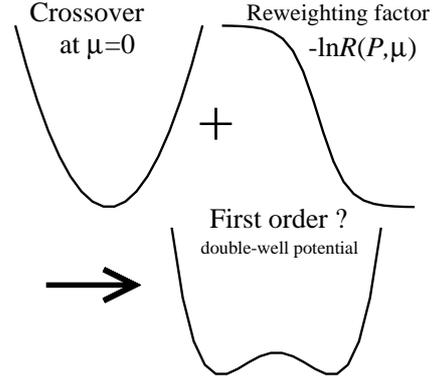}
}
\vspace{-5mm}
\caption{Schematic illustration of the effective potential and the reweighting factor. }
\label{fig:dwpot}
\end{figure}

The distribution function at $\mu \neq 0$ is 
$R(P; \mu) w(P;\beta,0)$, and thus the effective potential is defined by 
\begin{eqnarray}
V_{\rm eff}(P; \beta, \mu) & \equiv & -\ln [R(P; \mu) w(P; \beta,0)] 
\nonumber \\
&& \hspace{-20mm} = -\ln R(P; \mu) + V_{\rm eff}(P; \beta, 0).
\label{eq:potmu}
\end{eqnarray}
The shape of the effective potential can then be investigated 
at $\mu \neq 0$ once the reweighting factor $R(P; \mu)$ is obtained.
A schematic illustration of $V_{\rm eff}(P)$ is shown in Fig.~\ref{fig:dwpot}.

First, the peak position of the distribution function moves as $\mu$ changes,
which is determined by solving 
\begin{eqnarray}
\frac{\partial V_{\rm eff}}{\partial P} (P; \beta, \mu) 
= \frac{\partial V_{\rm eff}}{\partial P} (P; \beta, 0) 
- \frac{\partial (\ln R)}{\partial P} (P; \mu) =0. \ \ 
\label{eq:minmuq}
\end{eqnarray} 
Then, the effect from $\mu$ to the peak position is the same as that when $\beta$ (temperature) is changed. 
From the definition at $\mu=0$, 
the weight $w(P; \beta,0)$ and the effective potential becomes 
\begin{eqnarray}
w(P; \beta_{\rm eff},0) &=& e^{6 (\beta_{\rm eff} - \beta) N_{\rm site} P} w(P; \beta,0), \\
V_{\rm eff}(P; \beta_{\rm eff},0)
&=& V_{\rm eff}(P; \beta,0) -6 (\beta_{\rm eff} - \beta) N_{\rm site} P,
\label{eq:vrewbeta}
\end{eqnarray} 
under a change from $\beta$ to $\beta_{\rm eff}$. 
Hence, the change from $\beta$ to 
\begin{eqnarray}
\beta_{\rm eff}(\mu) 
\equiv \beta + (6N_{\rm site})^{-1} \partial (\ln R)/\partial P \ \
\label{eq:betaeff}
\end{eqnarray}
corresponds $(\beta, 0)$ to $(\beta, \mu)$ for the determination of the minimum of $V_{\rm eff} (P)$.
As we will see, the slope of $\ln R$ is positive. 
This explains why the phase transition temperature $(\beta)$ decreases when the density is increased.

Moreover, we find from Eq.~(\ref{eq:vrewbeta}) that 
the second derivative of $V_{\rm eff}(P)$ does not change under the change of $\beta$. 
This means that the curvature of $V_{\rm eff}(P)$ is independent of $\beta$. 
Thus, the critical value of $\mu$ can be estimated by measuring the curvature of 
the effective potential, 
\begin{eqnarray}
\frac{\partial^2 V_{\rm eff}}{\partial P^2} (P; \mu) 
= - \frac{\partial^2 (\ln R)}{\partial P^2} (P; \mu) 
+ \frac{\partial^2 V_{\rm eff}}{\partial P^2} (P; 0) =0, \ \ 
\end{eqnarray}
without fine-tuning $\beta$ to the critical value of $\beta$. 
Because the curvature of $V_{\rm eff}(P; \beta, 0)$ at $\mu=0$ is positive 
and the curvature of $V_{\rm eff}(P; \beta, \mu)$ at a second order phase 
transition point is zero, 
the parameter range where $-\ln R(P; \mu)$ has negative 
curvature is required for the existence of the critical point.

\begin{figure}[t]
\centerline{
\includegraphics[width=80mm]{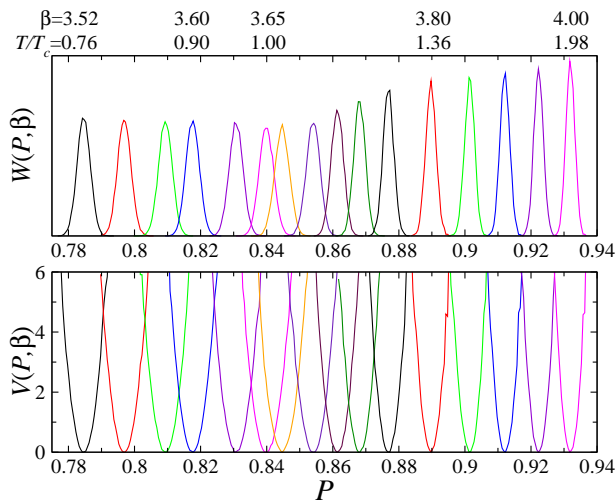}
}
\caption{Plaquette histogram and the effective potential in 2-flavor QCD at $\mu=0$ \cite{ejiri07,ejiri09}.}
\label{fig:plaq}
\end{figure} 

\begin{figure}[t]
\centerline{
\includegraphics[width=80mm]{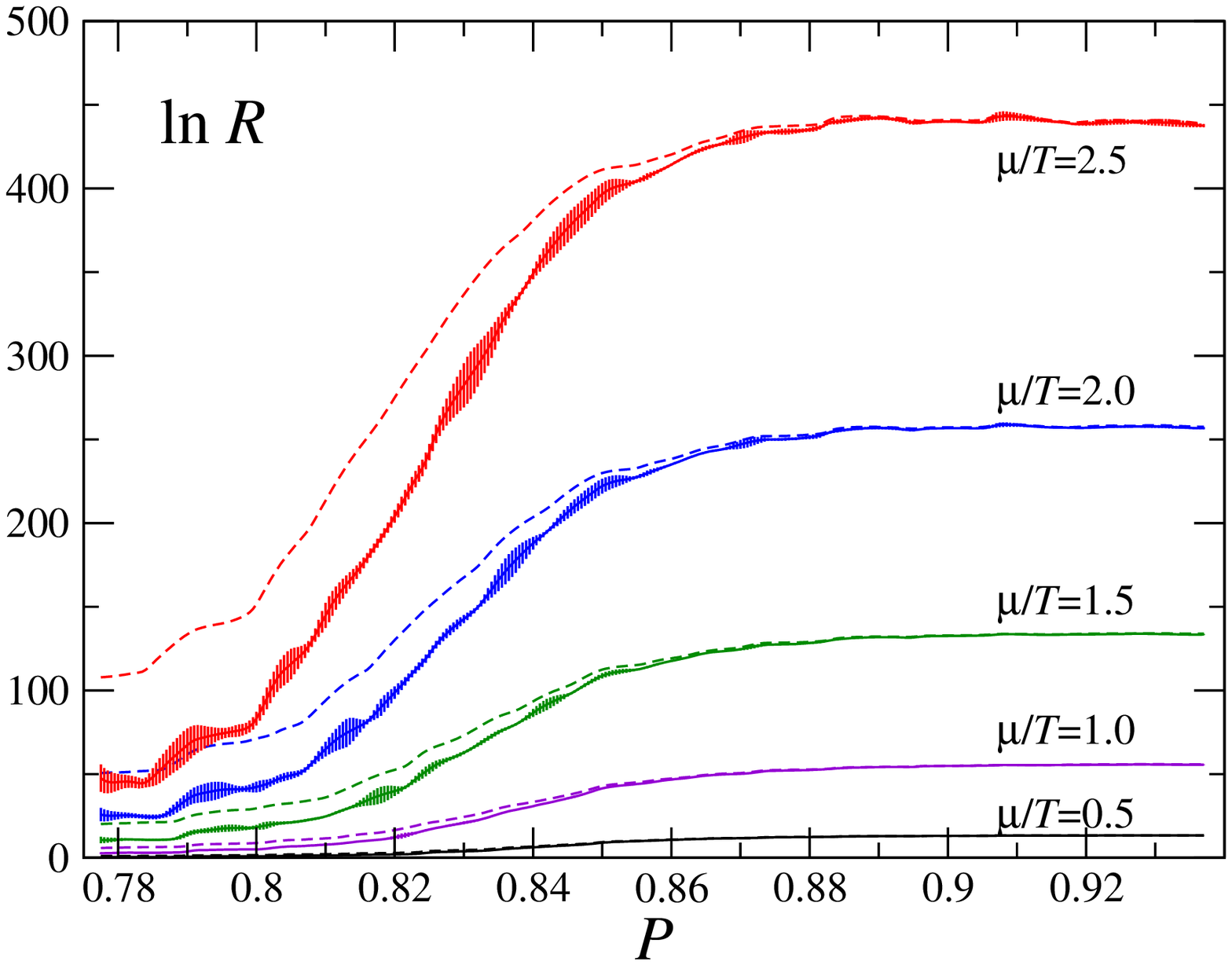}
}
\centerline{
\includegraphics[width=80mm]{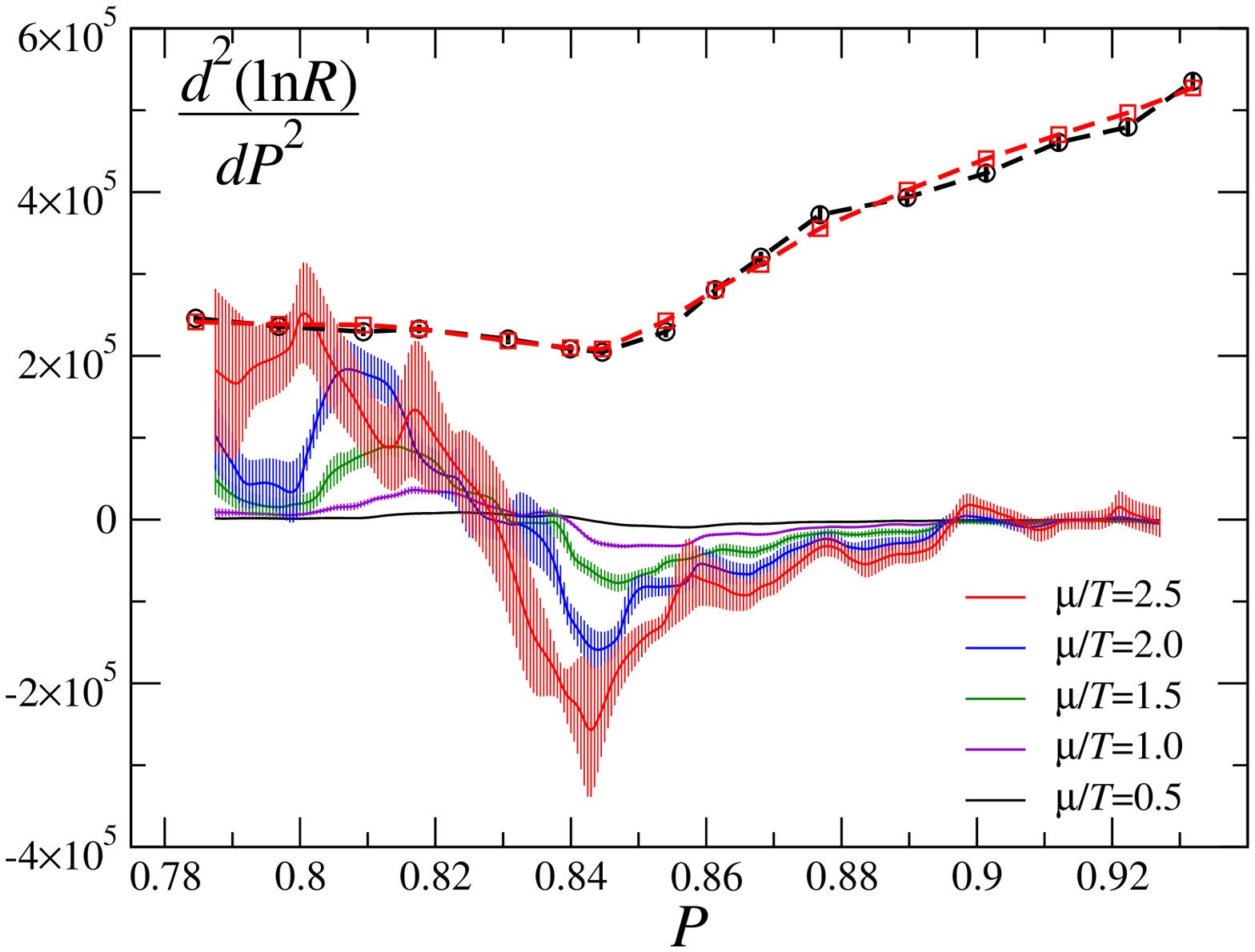}
}
\caption{Reweigiting factor (top) and its curvature (bottom) as functions of $P$ \cite{ejiri07}.
}
\label{fig:peff}
\end{figure}

The probability distribution function at non-zero $\mu$ has been 
calculated in \cite{ejiri07}
using the data obtained in \cite{BS05} with the 2-flavor p4-improved staggered quark action on a $16^3 \times 4$ lattice. 
The pion mass is $m_{\pi} \approx 770 {\rm MeV}$, 
which is heavier than the physical mass.
Further details on the simulation parameters are given in
Ref.~\cite{BS05,BS03}.
The distribution function $w(P)$ at $\mu=0$, i.e. the histogram of $P$, 
and the effective potential $V_{\rm eff}(P)$ are given in Fig.~\ref{fig:plaq} for each $\beta$. 
The values of $\beta$ and $T/T_c$ are shown above these figures. 
The potential $V_{\rm eff}(P)$ is normalized by the minimum value 
for each $\beta$.
Because the phase transition is a crossover transition for 2-flavor QCD 
with finite quark mass, 
the distribution function is always Gaussian type, 
i.e. the effective potential is always a quadratic function. 
The value of the plaquette at the potential minimum increases
as $\beta$ increases in accordance with the argument of 
the potential minimum.

The quark determinant $\det M(\mu) \equiv e^{\hat{F}} e^{i \hat{\theta}}$ is estimated using the data of the Taylor expansion coefficients in \cite{BS05};
\begin{eqnarray}
\hat{F} (\mu) 
& \equiv & N_{\rm f}\,  {\rm Re} \left[ \ln \left( \frac{\det M(\mu)}{\det M(0)} \right) \right]
\label{eq:teabs} \\
&& \hspace{-12mm} \approx N_{\rm f} \sum_{n=1}^{N_{\rm max}} \frac{1}{(2n)!} {\rm Re}
\left[ \frac{\partial^{2n} (\ln \det M(\mu))}{\partial (\mu/T)^{2n}} 
\right]_{\mu=0} \left( \frac{\mu}{T} \right)^{2n} ,
\nonumber \\
\hat\theta (\mu) 
& \equiv & N_{\rm f} \, {\rm Im} \,[\ln \det M(\mu)]
\label{eq:tatheta} \\
&& \hspace{-12mm} \approx N_{\rm f} \! \! \sum_{n=0}^{N_{\rm max}} \! \frac{1}{(2n-1)!} {\rm Im}
\left[ \frac{\partial^{2n-1} (\ln \det M(\mu))}{\partial (\mu/T)^{2n-1}} 
\right]_{\mu=0} \! \! \left( \frac{\mu}{T} \right)^{2n-1} .
\nonumber
\end{eqnarray}
Higher order terms than $\mu^6$ order are neglected. 
The truncation error has been estimated comparing 
the results up to $O(\mu^4)$ and $O(\mu^6)$ for $\mu/T \leq 2.5$ and 
is found to be small \cite{ejiri07}.
We use the delta function approximated by
$\delta(x) \approx 1/(\Delta \sqrt{\pi})$ $\exp[-(x/\Delta)^2]$, 
where $\Delta=0.0025$ is adopted.
Because the sign problem is serious for the calculation of $R(P; \mu)$, 
the method  discussed in Sec.~\ref{sec:phase} is used to avoid the sign problem;
\begin{eqnarray}
R(P,\mu) &=&
\left\langle \left| \frac{\det M(\mu)}{\det M(0)} \right|^{N_{\rm f}}
\right\rangle_{(P: {\rm fixed})} \left\langle e^{i \theta} \right\rangle_{(P, F: {\rm fixed})}
\nonumber \\
& \approx & \left\langle e^{F} \right\rangle_{(P: {\rm fixed})} 
\exp \left( -\frac{1}{2} \langle \hat{\theta}^2 \rangle_{(P: {\rm fixed})} \right) ,
\label{eq:rcum}
\end{eqnarray}
where the higher order terms are neglected, since the distribution 
of $\hat{\theta}$ is well approximated by a Gaussian function \cite{ejiri07}, 
and the contribution from the higher order terms seem to be small.
The reweighting factor $R(P; \mu)$ is plotted in Fig.~\ref{fig:peff} (top).
The dashed lines in Fig.~\ref{fig:peff} (top) are the results 
when the effect of the complex phase $e^{i \theta}$ is omitted. 
Because the contribution from the complex phase is not very large in this result, the error from the approximation to avoid the sign problem may be small.

To study the existence of a second order phase transition, 
we investigate the curvature of the potential.
Because the finite temperature transition is crossover for 2-flavor QCD with finite quark mass, the distribution function is always Gaussian, i.e. the effective potential is always a quadratic function. 
Assuming the Gaussian distribution, one can estimate the curvature of $V_{\rm eff}$ at $\mu=0$ from the relation between the plaquette susceptibility $\chi_P$ and the curvature of the potential, 
\begin{eqnarray}
\frac{d^2 V_0}{dP^2} = \frac{6N_{\rm site}}{\chi_P}.
\label{eq:curpot2}
\end{eqnarray}
where $V_0$ is the $V_{\rm eff}$ at $\mu=0$.
The slope of $V_0$ at $\mu=0$ can be also measured using 
Eq.~(\ref{eq:vrewbeta}) \cite{ejiri09}. 
When one performs a simulation at $\beta_0$, the slope is zero at the
minimum of $V_0(P;\beta_0)$, and the value of $P$ at the minimum
can be estimated by $\langle P \rangle_{\beta_0}$ approximately. 
Hence, we obtain $dV_0/dP$ at $\langle P \rangle_{\beta_0}$ by
\begin{equation}
\frac{dV_0}{dP} ( \langle P \rangle_{\beta_0}, \beta) 
= - 6(\beta - \beta_0)N_{\rm site}.
\label{eq:dveffdp}
\end{equation}
The dashed lines in Fig.~\ref{fig:peff} (bottom)
show the  curvature of the effective potential at $\mu=0$.
The circle symbols are $d^2V_0/dP^2$ determined by Eq.~(\ref{eq:curpot2}).
The square symbols are computed by the
numerical deferential of $dV_0/dP$ calculated by
Eq.~(\ref{eq:dveffdp}).
These results at $\mu=0$ obtained by two different methods are consistent with each other.

The second derivative $d^2 \ln R/dP^2$ is calculated by
fitting $\ln R$ to a quadratic function of $P$ with a range of
$P \pm 0.015$ and repeating with various $P$.
The result of $d^2 (\ln R)/dP^2 (P; \mu)$ is 
plotted as solid line in Fig.~\ref{fig:peff} (bottom). 
The magnitude of the curvature of $\ln R$ becomes larger 
as $\mu/T$ increases.
This figure indicates that the maximum value of 
$d^2 (\ln R)/dP^2 (P; \mu)$ at $P=0.80$ becomes larger 
than $d^2 V_0 /dP^2$ for $\mu/T \simge 2.5$. 
This means that the curvature of the effective potential 
$d^2 V_{\rm eff}/dP^2$
vanishes at $\mu/T \approx 2.5$ and a region of $P$ where the curvature is 
negative appears for large $\mu/T$, corresponding to a double-well potential. 

Further studies are, of course, necessary for the precise 
determination of the critical point in the $(T, \mu)$ plane, 
increasing the number of terms in the Taylor expansion of 
$\ln \det M$ and decreasing the quark mass in the simulation. 
However, this argument suggests the appearance of 
a first order phase transition line at large $\mu/T$.

\section{Phase diagram of $(2+N_{\rm f})$-flavor QCD}
\label{sec:manyflavor}

The ultimate goal of this study is to investigate the location of the critical surface 
in the light quark mass region of $(2+1)$-flavor QCD at finite density, 
including dynamical up, down and strange quarks.
However, recent lattice QCD studies suggest that the critical region at 
zero density is accessible only when the quark masses are very small   
and thus its determination may be difficult \cite{RBCBi09}.

In this section, we study QCD having two light flavors and many massive flavors \cite{yama12}.
As we will see, the first order transition region becomes wider as the number of massive 
flavors increases. Replacing the strange quark mass in Fig.~\ref{fig:massdep} 
(right) by the mass of many flavors, the boundary of first order comes to 
the upper left corner of the figure.
If the critical mass of QCD with many flavors is larger than that of 
$(2+1)$-flavor QCD, the boundary of the first order region can be 
investigated more easily for many-flavor QCD. 
Then, the many-flavor QCD can be a good testing ground for investigating 
$N_{\rm f}$-independent universal properties, such as the critical scaling 
near the tricritical point in Fig.~\ref{fig:massdep} (right), 
which is explained in appendix A.
This will provide important information for $(2+1)$-flavor QCD.

Moreover, the study of finite temperature many-flavor QCD is interesting for the construction of the Technicolor (TC) model built of many flavor QCD, i.e. vector-like SU(3) gauge theory with many fermions transforming as the fundamental representation.
In this model, the Higgs sector is replaced by a new strongly interacting gauge theory and its spontaneous chiral symmetry breaking causes electroweak (EW) symmetry breaking.
The EW baryogenesis scenario requires a strong first order phase transition.
As we discussed, the nature of the phase transition depends on the number of flavors and masses.
In realistic TC models, two flavors of them are exactly massless and the
mass of other $N_{\rm f}$ flavors must be larger than an appropriate lower
bound otherwise the chiral symmetry breaking produces too many (light pseudo)
Nambu-Goldstone (NG) bosons.
Three of them are absorbed into the longitudinal mode of the weak gauge
bosons, but any other NG bosons have not been observed yet.
On the other hand, the first order transition at small mass terminates
at the critical mass like $(2+1)$-flavor QCD.
Thus, if one requires the first order EW phase transition in TC model, 
it brings in the upper bound on the mass of $N_{\rm f}$ flavors.
This can be a motivation to study $(2+N_{\rm f})$-flavor QCD, too.

We consider QCD with two degenerate light quarks of the mass
$m_{\rm l}$ and the chemical potential $\mu$ and $N_{\rm f}$ heavy quarks.
Denoting the potential of 2-flavor QCD at $\mu =0$ by
$V_0(P; \beta)$, 
that of $(2+N_{\rm f})$-flavor QCD
is written as
\begin{eqnarray}
    V_{\rm eff}(P; \beta, m_f, \mu)
=   V_0(P; \beta_0)
  - \ln R(P; \beta, m_f, \mu; \beta_0), \ \ 
\label{eq:vefftrans}
\end{eqnarray}
with
\begin{eqnarray}
\ln R(P; \beta, m_f, \mu; \beta_0)
&=& 6(\beta - \beta_0)N_{\rm site}P 
\nonumber \\
& & \hspace{-38mm}
 + \ln
     \left\langle
     \displaystyle
     \left( \frac{\det M(m_{\rm l},\mu)}{\det M(m_{\rm l}   ,0)}
     \right)^{\!\! 2} \prod_{f=1}^{N_{\rm f}}
     \frac{\det M(m_f,  \mu_f)}{\det M(\infty,0)}
     \right\rangle_{\! (P: {\rm fixed})}, \ 
\label{eq:lnr}
\end{eqnarray}
where 
$ \langle \cdots \rangle_{(P: {\rm fixed})} \equiv 
\langle \delta(P- \hat{P}) \cdots \rangle_{\beta_0} /
\langle \delta(P- \hat{P}) \rangle_{\beta_0} $
and $\langle \cdots \rangle_{\beta_0}$ means the ensemble average over
2-flavor configurations generated at $\beta_0$, $m_{\rm l}$ and
vanishing $\mu$.
Since the $m_{\rm l}$ dependence is not discussed in the
following, $m_{\rm l}$ is omitted from the arguments.
$\beta_0$ is the simulation point, which may differ 
from $\beta$ in this method.

Restricting the calculation to the heavy quark region,
the second determinant for $N_{\rm f}$ flavors in Eq.~(\ref{eq:lnr}) is approximated
by the leading order of the hopping parameter expansion, which is given in Eq.~(\ref{eq:detM}) for the standard Wilson quark action.
For improved gauge actions such as 
$S_g = -6N_{\rm site} \beta [c_0 {\rm (plaquette)} 
+ c_1 {\rm (rectangle)}]$, 
additional $c_1\times O(\kappa^4)$ terms must be considered, 
where $c_1$ is the improvement coefficient and $c_0=1-8c_1$. 
However, since the improvement term does not affect the physics, 
we will cancel these terms by a shift of the coefficient $c_1$.

At a first order transition point, $V_{\rm eff}$ shows 
a double-well shape as a function of $P$, 
and equivalently the curvature of the potential $d^2 V_{\rm eff}/d^2P$ 
takes a negative value in a region of $P$.
To observe this behavior, $\beta$ must be adjusted to be the first 
order transition point.
However, as discussed in the previous section, 
$d^2V_{\rm eff}/dP^2$ is independent of $\beta$.
The fine tuning is not necessary.
Moreover, $d^2V_{\rm eff}/dP^2$ over the wide range of $P$ can be easily
obtained by combining data obtained at different $\beta$.
We therefore focus on the curvature of
the effective potential to identify the nature of the phase transition.

Denoting $h=2 N_{\rm f} (2\kappa_{\rm h})^{N_t}$ for $N_{\rm f}$ degenerate Wilson quarks with the hopping parameter $\kappa_{\rm h}$, 
or $h=N_{\rm f}/(4 \times (2m_{\rm h})^{N_t})$ for the staggered quarks with the mass $m_{\rm h}$, 
we obtain
$\ln R(P;\beta,\kappa_{\rm h},0;\beta_0)
=\ln\bar{R}(P; h,0)$
+(plaquette term) $+ O(\kappa_{\rm h}^{N_t+2})$
for $\mu=\mu_{\rm h}=0$ with
\begin{eqnarray}
\bar{R}(P; h,0)
= \left\langle \exp [6h N_s^3 \hat{\Omega}_{\rm R}] 
\right\rangle_{(P: {\rm fixed}, \beta_0)} .
\label{eq:rew2f}
\end{eqnarray}
$\bar{R}(P; h,0)$ is given by the Polyakov loop and is independent of $\beta_0$.
The plaquette term does not contribute to $d^2V_{\rm eff}/dP^2$
and can be absorbed by shifting  
$\beta \to \beta^{*} \equiv \beta + 48 N_{\rm f} \kappa_{\rm h}^4$ for Wilson quarks. 
Moreover, one can deal with the case with non-degenerate masses by adopting
$h=2 \sum_{f=1}^{N_{\rm f}} (2 \kappa_f)^{N_t}$ for the Wilson quark action or 
$h=(1/4)\sum_{f=1}^{N_{\rm f}} (2m_f)^{-N_t}$ for the staggered quark action.
Thus, the choice of the quark action is not important.
In the following, we discuss the mass dependence of $\bar{R}$ 
through the parameter $h$.

\subsection{Numerical results at zero density.}

\begin{figure}[t]
\centerline{
\includegraphics[width=80mm]{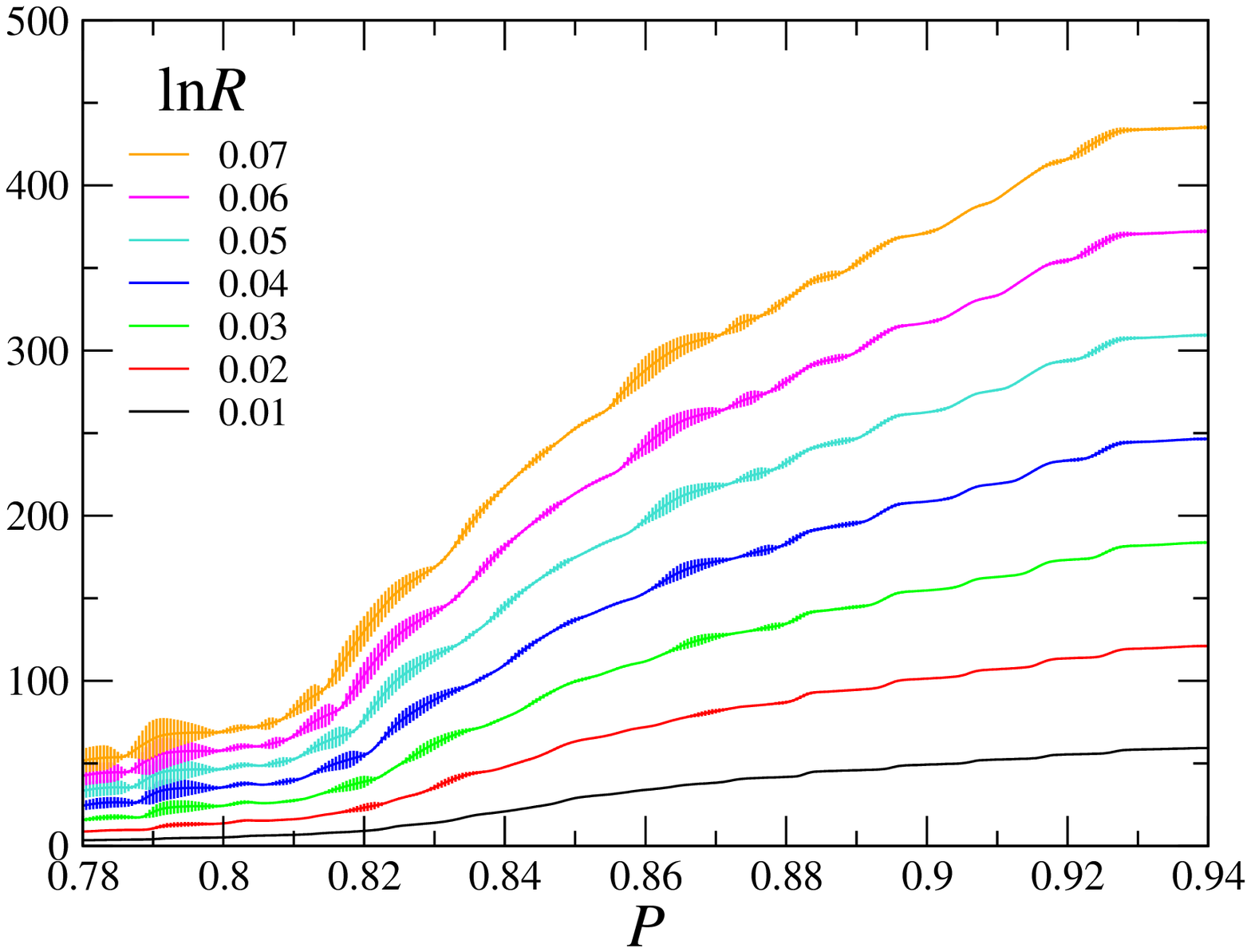}
}
\centerline{
\includegraphics[width=80mm]{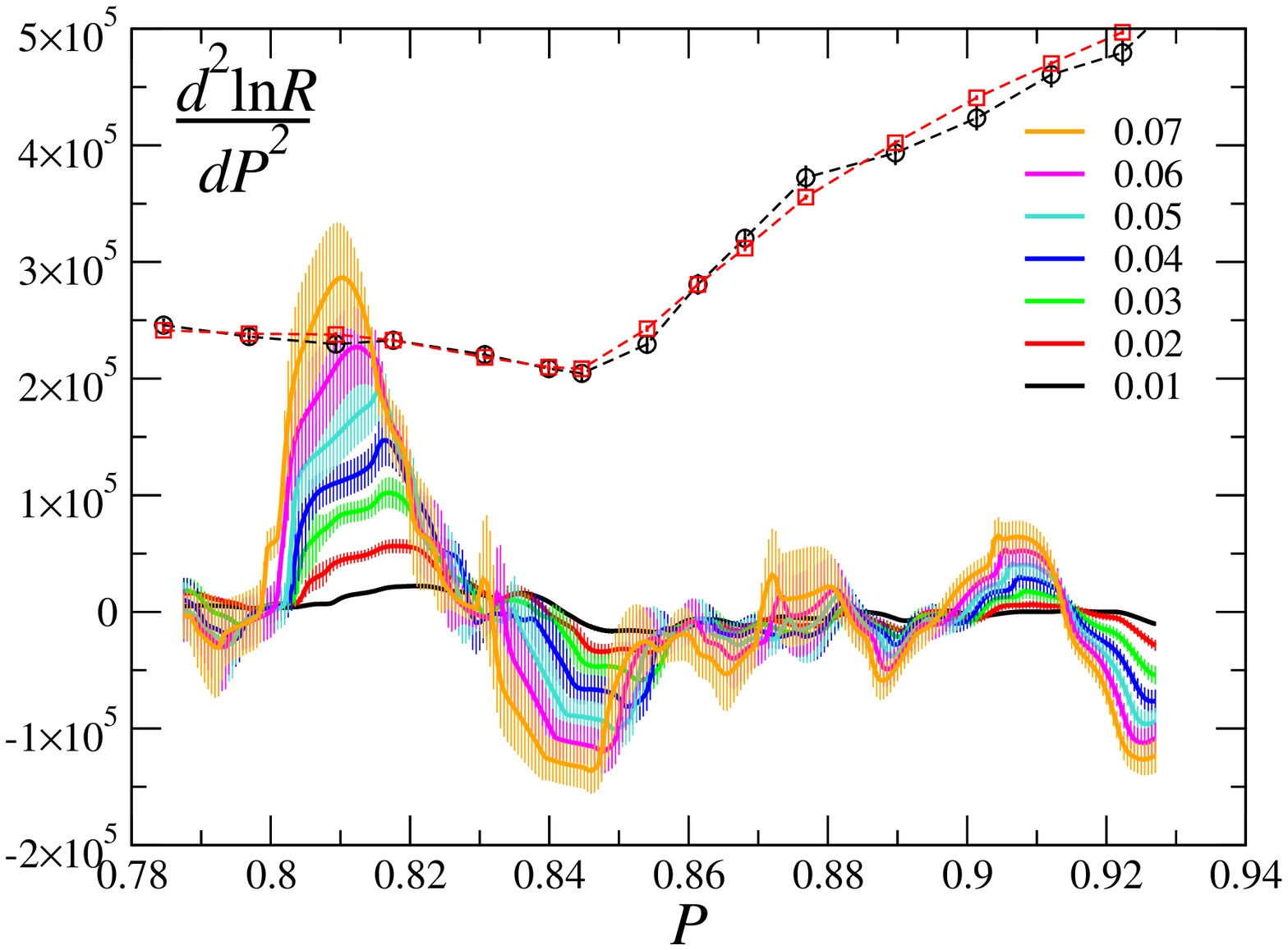}
}
\caption{{\em Top:} $\ln \bar{R} (P; h,0)$ as functions of the plaquette.
{\em Bottom:} The curvature of $\ln \bar{R} (P; h,0)$ 
for $h=0.01$ -- $0.07$ \cite{yama12}. 
The circle and square symbols are $d^2V_0/dP^2(P)$.}
\label{fig:lnr}
\end{figure} 

\begin{figure}[t]
\centerline{
\includegraphics[width=80mm]{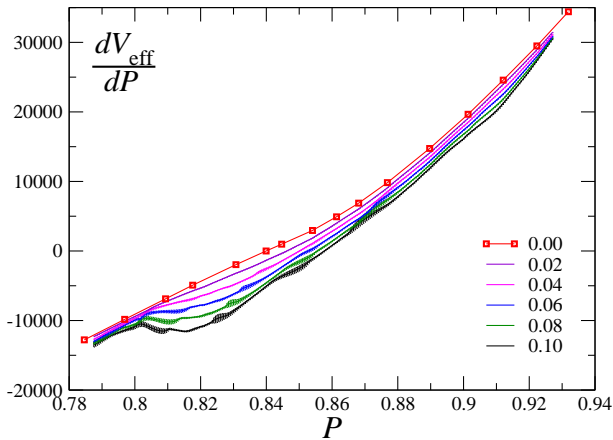}
}
\caption{The mass dependence of the 
slope of $V_{\rm eff} (P; \beta, h,0)$ 
normalized at $(\beta, h)=(3.65, 0)$ for $h=0.0$ -- $0.1$ \cite{yama12}.
The squares are $dV_0/d P$.}
\label{fig:vslp}
\end{figure}

We use the 2-flavor QCD configurations generated with the p4-improved
staggered quark and Symanzik-improved gauge actions \cite{BS05},
thus $\hat P=-S_g/(6N_{\rm site} \beta)$.
The same data set is used to study the phase structure of 2-flavor QCD
at finite density in Sec.~\ref{sec:2flavor}.
All configurations are used for the analysis at zero density, while the
finite density analysis is performed every 10 trajectories.
In the calculation of $\bar{R}(P; h,0)$, we use the delta
function approximated by
$\delta(x) \approx 1/(\Delta \sqrt{\pi})$ $\exp[-(x/\Delta)^2]$, 
where $\Delta=0.0025$ is adopted, again.
Because $\bar{R}(P; h,0)$ is independent of $\beta$, we mix all
data obtained at different $\beta$ as is done in the previous section.
The results for $\ln \bar R(P;h,0)$ are shown by solid lines in
the top panel of Fig.~\ref{fig:lnr} for $h=0.01$ -- $0.07$. 
A rapid increase is observed around $P \sim 0.82$, and 
the gradient becomes larger as $h$ increases.

The second derivative $d^2 \ln \bar R/dP^2$ is calculated by
fitting $\ln \bar R$ to a quadratic function of $P$ with a range of
$P \pm 0.015$ and repeating with various $P$.
The results are plotted in the bottom panel of Fig.~\ref{fig:lnr}, 
where $d^2V_0/dP^2$ in Fig.~\ref{fig:peff} (bottom) are plotted as the circles or the squares 
with dashed lines, again.
This figure shows that $d^2 (\ln \bar R)/dP^2$ becomes larger with
$h$, and the maximum around $P=0.81$ exceeds $d^2 V_0/dP^2$ for $h > 0.06$.
This indicates that the curvature of the effective potential,
\begin{eqnarray}
\frac{d^2 V_{\rm eff}}{dP^2} = \frac{d^2 V_0}{dP^2} - \frac{d^2 (\ln \bar R)}{dP^2}, 
\end{eqnarray}
vanishes at $h \approx 0.06$ and a region of $P$ where the curvature is negative
appears for large $h$.
We estimated the critical value $h_c$ at which the minimum of
$d^2 V_{\rm eff}/dP^2$ vanishes and obtained $h_c=0.0614(69)$.

To see the appearance of the first order transition in a different way,
we plot $dV_{\rm eff}/dP$ at finite $h$ for $\beta^*=3.65$ in
Fig.~\ref{fig:vslp}.
$dV_0/dP$ is computed by Eq.~(\ref{eq:dveffdp}).
The shape of $dV_{\rm eff}/dP$ is independent of $\beta$ because 
$d^2 V_{\rm eff}/dP^2$ is $\beta$-independent.
$dV_{\rm eff}/dP$ is monotonically increasing when $h$ is small,
indicating that the transition is crossover.
However, the shape of $dV_{\rm eff}/dP$ turns into an S-shaped function
at $h \approx 0.06$, corresponding to the double-well potential.

We have defined the parameter $h=2 N_{\rm f} (2\kappa_{\rm h})^{N_t}$ for the
Wilson quark.
Then, the critical $\kappa_{hc}$ corresponding $h_c$ decreases as
\begin{eqnarray}
\kappa_{hc}= \frac{1}{2} \left( \frac{h_c}{2N_{\rm f}} \right)^{1/N_t} 
\end{eqnarray}
with $N_{\rm f}$, and 
the truncation error from the higher order terms of the hopping parameter expansion 
in $\kappa_{\rm h}$ becomes smaller as $N_{\rm f}$ increases.
The application range of the hopping parameter expansion was 
examined in quenched QCD simulations with $N_t=4$, by explicitly
measuring the size of the next-to-leading order (NLO) terms of the
expansion~\cite{whot13}.
They found that the NLO contribution becomes comparable to that in the
leading order at $\kappa_{\rm h} \sim 0.18$.
Hence, this method may be applicable up to around $\kappa_{\rm h}\sim 0.1$.
For instance, in the case of $N_{\rm f}=10$ with $N_t=4$, 
$\kappa_{hc}$ is 0.118.

\subsection{Numerical results at finite density.}

\begin{figure}[tb]
\centerline{
\includegraphics[width=43mm]{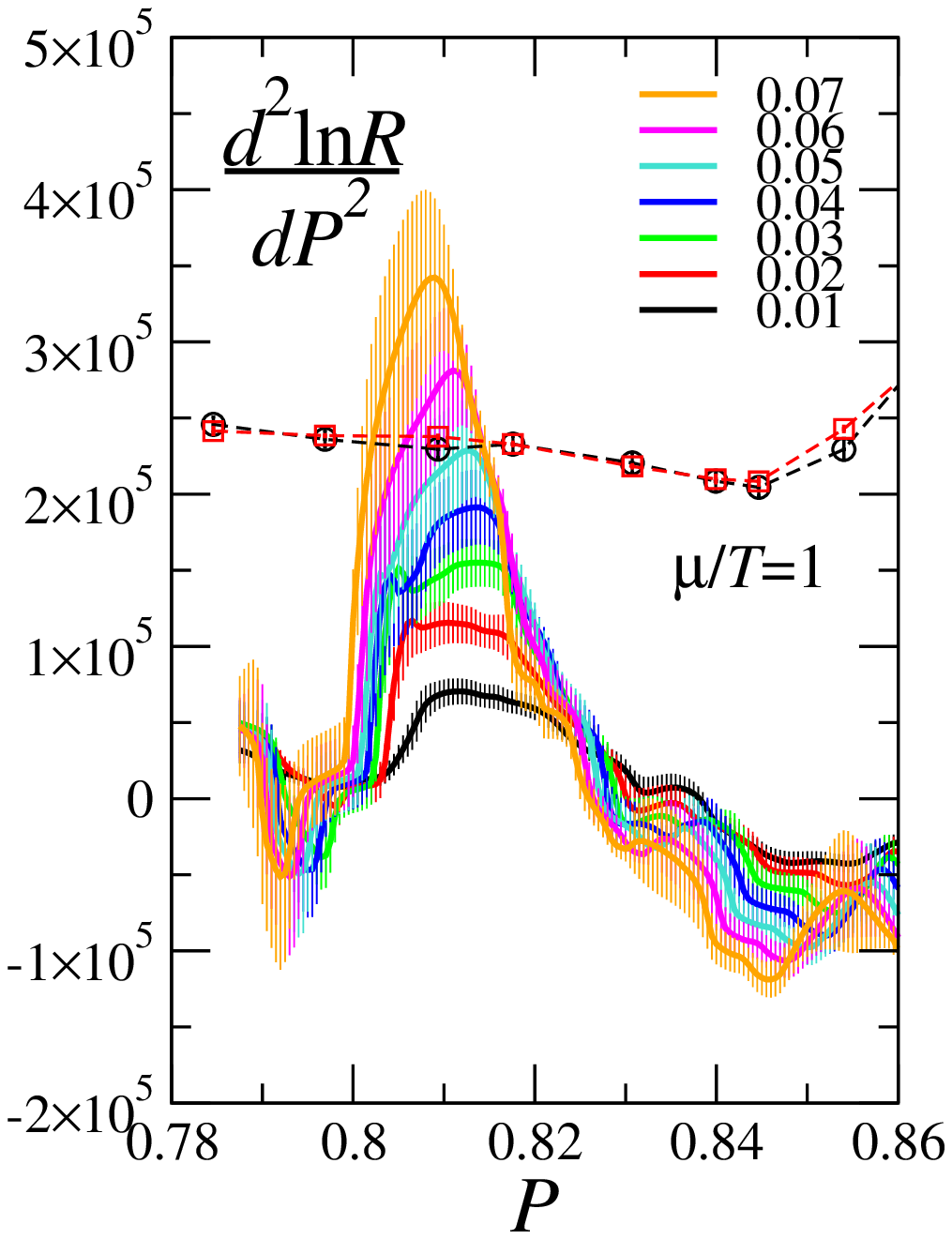}
\includegraphics[width=43mm]{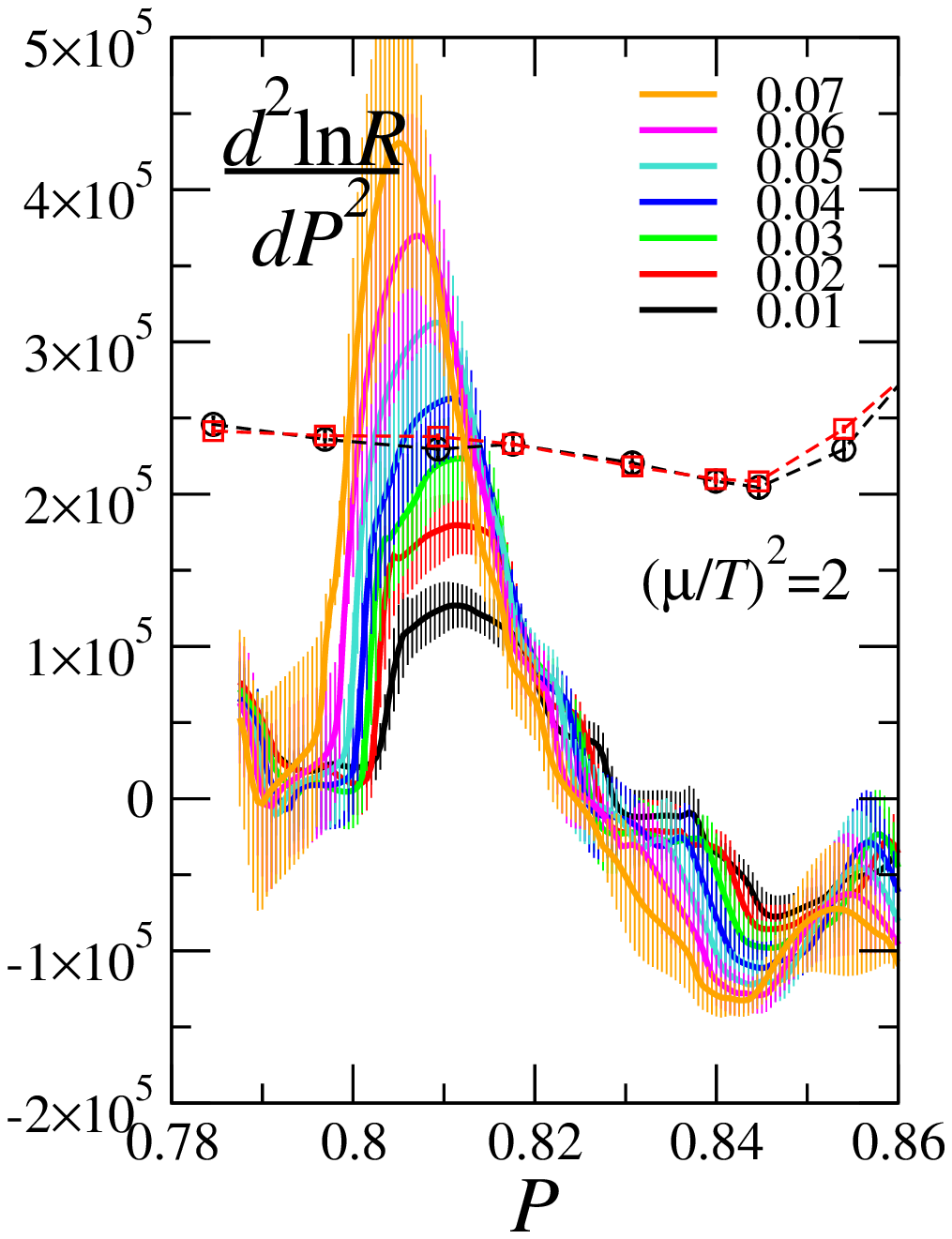}
}
\caption{The curvature of $\ln \bar{R} (P; h, \mu)$ 
as a function of the plaquette at $\mu /T=1.0$ (left) and $\sqrt{2}$ (right) 
when $\mu_{\rm h}=0$.
}
\label{fig:curmu}
\end{figure}

\begin{figure}[t]
\centerline{
\includegraphics[width=80mm]{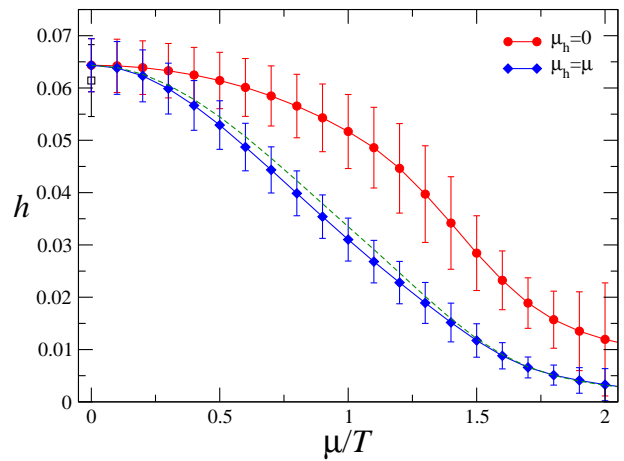}
}
\caption{The critical line in the $(h, \mu)$ plane for 
$\mu_{\rm h}=0$ (circles) and for $\mu_{\rm h}=\mu$ (diamonds) \cite{yama12}. 
In the region above this line, the transition is of first order. 
The square at $\mu=0$ is computed using all configurations, 
and the others are measured with every 10.}
\label{fig:crtmu}
\end{figure}

Next, we turn on a chemical potential $\mu$ for two light quarks and 
$\mu_{\rm h}$ for $N_{\rm f}$ flavors, and 
discuss the $\mu$ dependence of the critical mass.
As discussed above, we can investigate the critical region more easily for 
large $N_{\rm f}$.
$\bar R(P;h,\mu)$ is then given by  
$\langle (\det M(m_{\rm l},\mu)/ \det M(m_{\rm l},0))^2$ $\times$ 
$(\det M(m_{\rm h},\mu_{\rm h})/ \det M(\infty,0))^{N_{\rm f}} \rangle_{(P: {\rm fixed})}$.
The quark determinant is computed using the Taylor expansion of \\ 
$\ln [\det M(m_{\rm l}, \mu)/ \det M(m_{\rm l}, 0)]$ in terms of 
$\mu/T$ up to $O[\mu^6]$ and the Gaussian
approximation is applied to avoid the sign problem as explained in
Sec.~\ref{sec:2flavor}.
This approximation is valid for small $\mu$.
Figures \ref{fig:curmu} show the curvatures of $V_0$ and
$\ln \bar R$ at $\mu/T=1$ (left) and $\sqrt{2}$ (right) with $\mu_{\rm h}=0$. 
The maximum values of $d^2 \ln \bar R/ dP^2$ increases with $\mu$.
This means the critical $h$ is smaller at finite $\mu$. 

The circle symbols in Fig.~\ref{fig:crtmu} are the critical value of $h$ 
as a function of $\mu$ for $\mu_{\rm h}=0$, the diamond symbols are those 
for $\mu_{\rm h}=\mu$.
The critical $h$ $(h_c)$ for the case of $0 < \mu_{\rm h} < \mu$ can be estimated by the following way.
The imaginary part of the heavy quark determinant can be neglected when $\mu_{\rm h}$ is small or $\mu_{\rm h}$ is large ($h$ is small). 
Then, the difference between $\mu_{\rm h}=0$ and finite $\mu_{\rm h}$ is just the factor $\cosh(\mu_{\rm h}/T)$ in front of $h$, as discussed in Sec.~\ref{sec:heavy}. 
Once we find $h_c$ in the case of $\mu_{\rm h}=0$, $h_c (\mu_{\rm h})$ must be given by the following equation; 
$h_c (\mu_{\rm h}) \cosh(\mu_{\rm h}/T) = h_c (\mu_{\rm h}=0)$.
The dashed line is $h_c (\mu_{\rm h}=0)/ \cosh(\mu/T)$.
The results of $h_c$ at $\mu_{\rm h}=\mu$ satisfy this relation for large $\mu$ and small $\mu$. 
In the region above this critical line, the effective potential has the negative
curvature region, indicating the transition is of first order.
It is clear that the first order region becomes wider as $\mu$ increases.
If the same behavior is observed in $(2+1)$-flavor QCD, 
this gives the strong evidence for the existence of the critical point at
finite density in the real world.

\subsection{Towards $(2+1)$-flavor QCD.}

Although this analysis is valid only for large $N_{\rm f}$, it gives a frame of
reference for the study of critical mass at finite $\mu$.
Notice that $\ln \bar{R}(P;h,\mu)$ is given by the sum of
$\ln \bar R(P;0,\mu)$ and $\ln \bar R(P;h,0)$ approximately and that
the behavior of $\ln \bar{R}(P;h,0)$ in Fig.~\ref{fig:lnr} is very similar 
to that of $\ln R(P;0,\mu)$ in Fig.~\ref{fig:peff}.
$\ln R(P;0,\mu)$ is estimated from the quark number susceptibility
at small $\mu$ and $\ln \bar{R}(P;h,0)$ is obtained from the Polyakov
loop at small $\kappa_{\rm h}$.
Both the quark number susceptibility and the Polyakov loop rapidly
increase at the same value of $P$ near the transition point, which
enhances the curvature of $\ln R$.
Therefore, the critical $h$ decreases with $\mu$ or 
the critical $\mu$ decreases with $h$.
The same argument is possible for $(2+1)$-flavor QCD. 

Moreover, an interesting application is to study 
universal scaling behavior near the tricritical point, which is explained in appendix A.
If the chiral phase transition in the two flavor massless limit is
of second order, the boundary of the first order
transition region $m_{\rm l}^c (m_{\rm h})$ is expected to behave as 
$m_{\rm l}^c \sim |m_{\rm E} -m_h|^{5/2}$
in the vicinity of the tricritical point, 
$(m_{\rm l}, m_{\rm h}, \mu)=(0, m_{\rm E}, 0)$,
from the mean field analysis.
This power behavior is universal for any $N_{\rm f}$. 
The density dependence is important as well, which is expected to be
$m_{\rm l}^c \sim |\mu|^5$.
Starting from large $N_{\rm f}$, the systematic study of properties of 
QCD phase transition would be possible.

\section{Canonical approach}
\label{sec:canonical}

\subsection{Effective potential of the quark number}

Another interesting approach is to construct the canonical 
partition function ${\cal Z}_{\rm C} (T,N)$ by fixing the total 
quark number $(N)$ or quark number density $(\rho)$, i.e. 
the baryon number or baryon number density. 
Using the canonical partition function, one can also discuss the effective potential as a function of the quark number.
In this section, we denote the grand partition function as 
${\cal Z}_{\rm GC}(T,\mu)$ to distinguish it from the canonical partition function 
${\cal Z}_{\rm C} (T,N)$ explicitly. 
The relation between ${\cal Z}_{\rm GC}(T,\mu)$ and 
${\cal Z}_{\rm C} (T,N)$ is given by 
\begin{eqnarray}
{\cal Z}_{\rm GC}(T,\mu) 
&=& \int {\cal D}U \left( \det M(\mu/T) \right)^{N_{\rm f}} e^{-S_g}
\nonumber \\
&=& \sum_{N} \ {\cal Z}_{\rm C}(T,N) e^{N \mu/T}. 
\label{eq:cpartition} 
\end{eqnarray}
Because this equation is a Laplace transformation from ${\cal Z}_{\rm C}$ to ${\cal Z}_{\rm GC}$ essentially, the canonical partition function is obtained from ${\cal Z}_{\rm GC} (T,\mu)$ by an inverse Laplace transformation.

In order to investigate the net quark number giving the largest 
contribution to ${\cal Z}_{\rm GC} (T,\mu)$, it is worth introducing an effective potential $V_{\rm eff}$ as 
a function of $N$, 
\begin{eqnarray}
V_{\rm eff}(N,T,\mu) 
& \equiv & - \ln {\cal Z}_{\rm C}(T,N) -N \frac{\mu}{T} 
= \frac{f(T,N)}{T} -N \frac{\mu}{T} , 
\nonumber \\
{\cal Z}_{\rm GC}(T,\mu) 
&=& \sum_{N} \ e^{-V_{\rm eff}}, 
\label{eq:effp} 
\end{eqnarray}
where $f$ is the Helmholtz free energy. 
Using the effective potential of the net quark number, the nature of phase transition can be studied.

If there is a first order phase transition region, 
we expect that this effective potential has minima at more than one value 
of $N$. At the minima, the derivative of $V_{\rm eff}$ satisfies 
\begin{eqnarray}
\frac{\partial V_{\rm eff}}{\partial N} (N, T, \mu ) 
= -\frac{\partial (\ln {\cal Z}_{\rm C})}{\partial N} (T,N) 
- \frac{\mu}{T} =0 .
\label{eq:dpotdn} 
\end{eqnarray}
Hence, in the first order transition region of $T$, we expect 
$\partial (\ln {\cal Z}_{\rm C})/ \partial N (T,N) \equiv - \mu_q^*/T$ 
takes the same value at different $N$. 
Here, $\mu_q^* (T,N)$ is the chemical potential which gives a minimum 
of the effective potential at $(T, N)$ and becomes $\mu$ in the thermodynamic limit because the potential is minimized in the large volume limit.

\begin{figure}[t]
\centerline{
\includegraphics[width=85mm]{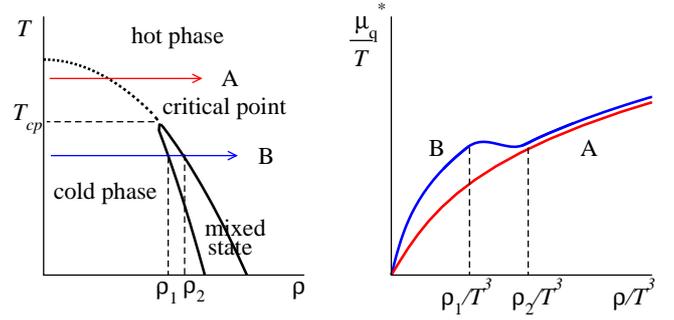}
}
\caption{Phase structure in the $(T, \rho)$ plane and the behavior of 
$\mu_q^*/T$ as a function of $\rho$.}
\label{fig:cano}
\end{figure} 

The phase structure in the $(T, \rho)$ plane and the expected behavior 
of $\mu_q^*/T$ are sketched in the left and right panels of 
Fig.~\ref{fig:cano}, respectively. 
The bold lines in the left figure are the phase transition line. 
We expect that the transition is crossover at low density and becomes 
first order at high density. 
Since two states coexist on the first order transition line, the phase 
transition line splits into two lines in the high density region, 
and the two states are mixed in the region between two lines. 
The expected behavior of $\mu_q^*$ along the lines A and B are shown 
in the right figure. 
When the temperature is higher than the temperature at the critical 
point $T_{cp}$ (line A), 
$\mu_q^*$ increases monotonically as the density increases. However, 
for the case below $T_{cp}$ (line B), this line crosses the mixed state. 
Because the two states of $\rho_1$ and $\rho_2$ are realized at 
the same time, $\mu_q^*$ does not increase in this region between 
$\rho_1$ and $\rho_2$. 

The Glasgow method \cite{Gibbs86,Hase92,Bar97,Bar98,Krat05,Krat06} has been a well-known method to 
compute the canonical partition function for the standard staggered fermion, 
and the same technique has been developed by \cite{nakamura10,alex11} for Wilson fermions.
Moreover, a method to perform simulations with canonical ensemble directly has been proposed 
in \cite{Alex05,Li10,Li11}.
In this paper, we focus on a method based on a saddle point approximation \cite{ejiri08} 
to illustrate the canonical approach. 
Reducing the computational cost by the approximation,
the first order like behavior is observed in 2-flavor QCD.

\subsection{Inverse Laplace transformation by Saddle point approximation}

We demonstrate the calculation of the quark number effective potential 
in 2-flavor QCD \cite{ejiri08}.
From Eq.~(\ref{eq:cpartition}),
the canonical partition function can be obtained by an inverse Laplace 
transformation \cite{Hase92};
\begin{eqnarray}
&& \hspace{-3mm} {\cal Z}_{\rm C}(T,N) \nonumber \\
&=& \frac{3}{2 \pi} \int_{-\pi/3}^{\pi/3} \!
e^{-N (\mu_0/T+i\mu_i/T)} {\cal Z}_{\rm GC}(T, \mu_0+i\mu_i) 
d \! \left( \! \frac{\mu_i}{T} \!\right), \hspace{5mm}
\label{eq:canonicalP} 
\end{eqnarray}
where $\mu_0$ is an appropriate real constant and $\mu_i$ is 
a real variable. Note that 
${\cal Z}_{\rm GC}(T, \mu +2\pi iT/3) = {\cal Z}_{\rm GC}(T, \mu)$
\cite{Rob86}.
The grand canonical partition function can be evaluated by 
the calculation of the following expectation value at $\mu=0$.
\begin{eqnarray}
\frac{{\cal Z}_{\rm GC}(T, \mu)}{{\cal Z}_{\rm GC}(T,0)}
&=& \frac{1}{{\cal Z}_{\rm GC}} \int {\cal D}U 
\left( \frac{\det M(\mu/T)}{\det M(0)} \right)^{N_{\rm f}} 
\nonumber \\
&& \hspace{-20mm} \times 
\det M(0) ^{N_{\rm f}} e^{-S_g} 
= \left\langle 
\left( \frac{\det M(\mu/T)}{\det M(0)} \right)^{N_{\rm f}}
\right\rangle_{(T, \mu=0)} . \ \ \ 
\label{eq:normZGC} 
\end{eqnarray}

We then perform the integral in Eq.~(\ref{eq:canonicalP}).
However, the calculation of the quark determinant needs much computational cost usually.
To reduce the computational cost, we use a saddle point approximation. 
If one selects a saddle point as $\mu_0$ in Eq.~(\ref{eq:canonicalP}). 
The information which is needed for the integral is only around the saddle point when the volume is sufficiently large. 
Moreover, if we restrict ourselves to study the low density region, the value of $\det M (\mu/T)$ near the saddle point can be estimated by the Taylor expansion around $\mu=0$. 
The calculations by the Taylor expansion are much cheaper than the exact calculations and the studies using large lattices are possible. 
Also, the truncation error can be systematically controlled by increasing the number of the expansion coefficients.
Therefore, we use a saddle point approximation.

We denote the quark number density in a lattice unit and physical unit as
$\bar{\rho}=N/N_s^3$ and $\rho/T^3=\bar{\rho} N_t^3$, respectively. 
We assume that a saddle point $z_0$ exists in the complex $\mu/T$ 
plane for each configuration, which satisfies
$D'(z_0) - \bar{\rho} =0$, 
where $(\det M(z)/\det M(0))^{N_{\rm f}} = \exp[N_s^3 D(z)] $ and 
$D'(z)=dD(z)/dz$.
We then perform a Taylor expansion around the saddle point and obtain 
the canonical partition function, 
\begin{eqnarray}
\frac{{\cal Z}_{\rm C}(T, \bar{\rho} V)}{{\cal Z}_{\rm GC}(T,0)} 
& \approx & \frac{3}{2 \pi} 
\left\langle \int_{-\pi/3}^{\pi/3} \! \!
e^{-N (z_0+ix)} e^{N_s^3 D(z_0+ix)}
dx \right\rangle_{ \! \!(T, \mu=0)} \nonumber \\
&& \hspace{-18mm} \approx \frac{3}{2 \pi} 
\left\langle \int_{-\pi/3}^{\pi/3} e^{V \left(D (z_0) 
- \bar{\rho} z_0 - \frac{1}{2} D'' (z_0) x^2 + \cdots \right)} 
dx \right\rangle_{ \! (T, \mu=0)} \nonumber \\
&& \hspace{-18mm} \approx \frac{3}{\sqrt{2 \pi}} 
\left\langle e^{V \left( D(z_0) - \bar{\rho} z_0 \right)} 
e^{-i \alpha/2} \frac{1}{\sqrt{V |D''(z_0)|}} \right\rangle_{\! (T, \mu=0)}.
\label{eq:zcspa}
\end{eqnarray}
Here, $ D''(z) = d^2 D(z) /dz^2 \equiv |D''(z)| e^{i \alpha},$ 
and $V \equiv N_s^3$.
We chose a path which passes through the saddle point $z_0$ in the complex $\mu/T$ plane.
Higher order terms in the expansion of $D(z)$ become negligible when 
the volume $V$ is sufficiently large. 

We calculate the derivative of the effective potential with respect to $N$ or $\rho$.
Within the framework of the saddle point approximation, 
this quantity can be evaluated by 
\begin{eqnarray}
\frac{\mu_q^*}{T} 
&=& - \frac{1}{V} \frac{\partial \ln {\cal Z}_C (T, \bar{\rho} V)}
{\partial \bar{\rho}}
\nonumber \\
&& \hspace{-7mm} \approx \frac{
\left\langle z_0 \ e^{V \left( D(z_0)
- \bar{\rho} z_0 \right) } 
e^{-i \alpha /2} \sqrt{ \frac{1}{V |D''(z_0)|}}
\right\rangle_{(T, \mu=0)}}{
\left\langle e^{V \left( D(z_0) 
- \bar{\rho} z_0 \right) } 
e^{-i \alpha /2} \sqrt{ \frac{1}{V |D''(z_0)|}}
\right\rangle_{(T, \mu=0)}}. 
\label{eq:chemap}
\end{eqnarray}
This equation is similar to the formula of the reweighting method 
for finite density. 
The operator in the denominator corresponds to a reweighting factor, 
and $\mu_q^* /T$ is an expectation value of the saddle 
point calculated with this modification factor.

\begin{figure}[t]
\centerline{
\includegraphics[width=80mm]{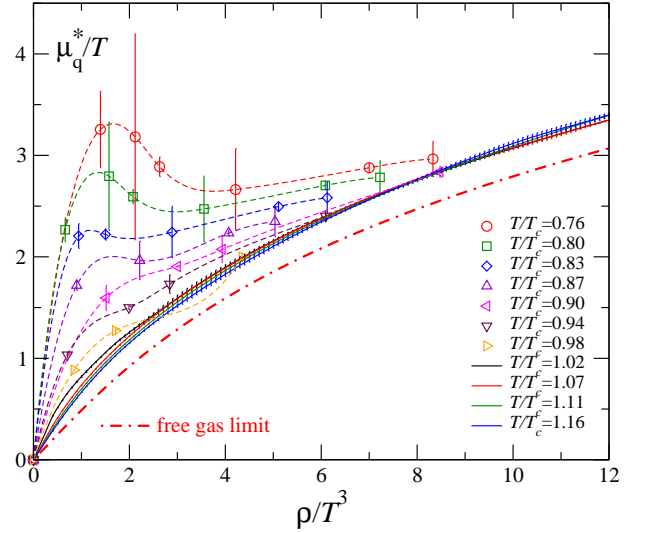}
}
\caption{Chemical potential vs. quark number density for $N_f=2$ with a saddle point approximation \cite{ejiri08}.
\label{fig:chem}
}
\end{figure}

We compute the derivative of $\ln {\cal Z}_C$ \cite{ejiri08} using the data obtained in \cite{BS05} with the 2-flavor p4-improved staggered quark action, $m_{\pi} \approx 770 {\rm MeV}$. 
Because the modification factor is a complex number, this calculation suffers from the sign problem. 
To eliminate the sign problem, the Gaussian approximation discussed in Sec.~\ref{sec:phase} is used. 
If one assumes that the distribution of the complex phase is well-approximated by a Gaussian function, the complex phase factor $e^{i \hat{\theta}}$ can be replaced by $\exp[- \langle \hat{\theta}^2 \rangle /2]$. 

The quark determinant is estimated by the Taylor expansion around $\mu=0$ up to $O(\mu^6)$. 
We find a saddle point $z_0$ in the complex $\mu/T$ plane for each configuration numerically, calculating $D'(\mu/T) - \bar{\rho}$ from the Taylor expansion coefficients. 
Because the calculation of Eq.~(\ref{eq:chemap}) is similar to the reweighting method, the important configurations will change by the modification factor. 
To avoid this problem, the multi-point reweighting method is used. 
The important configurations are thus automatically selected among all configurations generated at some $\beta$.

The result of $\mu_q^*/T$ is shown in Fig.~\ref{fig:chem} as a function of $\rho/T^3$ for each temperature $T/T_c$ $(\beta)$. Here, $T_c$ is the transition temperature at $\mu=0$.
The dot-dashed line is the value of the free quark-gluon gas in 
the continuum theory;
\begin{eqnarray}
\frac{\rho}{T^3} = N_{\rm f} \left[ \frac{\mu}{T} 
+ \frac{1}{\pi^2} \left( \frac{\mu}{T} \right)^3 \right].
\end{eqnarray}
From this figure, we find that a qualitative feature of $\mu_q^*/T$ 
changes around $T/T_c \sim 0.8$, i.e. $\mu_q^*/T$ increases monotonically 
as $\rho$ increases above 0.8, whereas it shows an S-shape below 0.8. 
This means that there is more than one value of $\rho/T^3$ for 
one value of $\mu_q^*/T$ below $T/T_c \sim 0.8$.
This is a signature of a first order phase transition. 
Although the approximation is used, the critical value of $\mu_q^*/T$ is about $2.4$, which is roughly consistent with the critical point estimated in Sec.~\ref{sec:2flavor} by calculating the plaquette effective potential using the same configurations, $(T/T_c, \mu/T) \approx (0.76, 2.5)$. 
The difference between these two results may be a systematic error. 
Further studies are necessary to predict the critical point quantitatively, 
but the distribution function of the baryon number relates directly to 
the baryon number fluctuation in the heavy-ion collisions, and
we find that the canonical approach is useful to study the phase structure at finite density

\section{Summary}
\label{sec:summary}

We summarized a series of studies about the QCD phase transition using the probability distribution functions.
The quark mass dependence and chemical potential dependence of the nature of the phase transition is investigated. 
In the heavy quark region, we evaluated the quark determinant by the hopping parameter expansion, and calculated the probability distribution function. Through the shape of the distribution function, the critical surface separating the first order transition and crossover regions was found in the quark masses and chemical potential parameter space, which is shown in Fig.~\ref{fig:crtsur}.

The existence of the critical point at finite density is discussed in 2-flavor QCD with an intermediate quark mass. 
We estimated $\ln \det M$ from the data of a Taylor expansion up to $O(\mu^6)$ and 
assumed the distribution of the complex phase of $\det M$ to be Gaussian. 
Then, it is found that the shape of the effective potential which is a quadratic function at 
$\mu=0$ changes to a double-well type at large $\mu/T$. 
Although further investigations must be needed, this argument strongly suggests 
the existence of the first order phase transition line in the $(T, \mu)$ plane.

Moreover, we studied $(2+N_{\rm f})$-flavor QCD, where the mass of two flavors is fixed 
to be small and the others are large, and determined the critical mass of 
the $N_{\rm f}$ massive quarks at which the first order transition changes to crossover.
The critical mass is found to become larger as $N_{\rm f}$ increases.
Furthermore, the chemical potential dependence of the critical mass
is investigated for large $N_{\rm f}$, and the critical mass is found to
increase with $\mu$. 
The study of the large flavor QCD may be a good starting point 
for the systematic study of the properties of QCD phase transition.

Finally, we made a comment on the probability distribution function of 
net baryon number (or net quark number) and 
the baryon number fluctuations in heavy-ion collisions.

\section*{Acknowledgments}
We would like to thank the members of the WHOT-QCD collaboration, N. Yamada 
and H. Yoneyama for discussions and collaborations.
This work is supported by Grants-in-Aid of the Japanese Ministry of Education, Culture, Sports, Science and Technology (No.\ 23540295) 
and by the Grant-in-Aid for Scientific Research on Innovative Areas (No.\ 23105706).

\section*{Appendix A: Mean field argument near the tricritical point}
\label{sec:appA}

We discuss the tricritical point by a mean field analysis of the standard sigma model. The tricritical point is located on the $m_{ud}=0$ axis in Fig.~\ref{fig:massdep} (right), at which the first order transition changes to second order. 
In the vicinity of the tricritical point at $\mu=0$, the effective potential in terms of the chiral order parameter $\sigma$ is modeled by the following equation; 
\begin{eqnarray}
V_{\rm eff} (\sigma) = \frac{1}{2} a \sigma^2 + \frac{1}{4} b \sigma^4 
+ \frac{1}{6} c \sigma^6 -h \sigma, 
\end{eqnarray}
where we assume $c>0$ so that $V_{\rm eff}$ is bounded from below for large $|\sigma|$. 
For the case of two light quarks with the mass $m_{ud}$ and a massive quark with the mass $m_s$, 
the coefficients, $a, b$ and $h$ may be parameterized as 
\begin{eqnarray}
&& \! a= a_t t + a_s s + a_{\mu} \mu^2, \hspace{3mm} 
b= b_s s + b_{\mu} \mu^2, \hspace{3mm} 
h= m_{ud} , \ \ \ \\
&& \! t= \frac{T-T_E}{T_E}, \hspace{3mm} 
s= \frac{m_E -m_s}{m_E},
\label{eq:sigmapara}
\end{eqnarray}
where $(T_E, m_E)$ is $(T, m_s)$ at the tricritical point. 
The coefficient $b$ controls the order of phase transition. 
Assuming a symmetry under $\mu$ to $-\mu$, the leading contribution to $b$ must be $\mu^2$ at low density. 

Since the effective potential is $O(\sigma^4)$ on the second order critical surface, 
\begin{eqnarray}
\frac{\partial^n V_{\rm eff}}{\partial \sigma^n} = 0. \hspace{5mm}
(n=1,2,3)
\end{eqnarray}
Solving these equations, we obtain 
\begin{eqnarray}
h= \frac{8c}{3} \left( \frac{a}{5c} \right)^{5/4} 
{\rm and} \
h= \frac{8c}{3} \left( \frac{-3b}{10c} \right)^{5/2} \! .  
(a \geq 0, b \leq 0) \ \ \ \ 
\end{eqnarray}
From the first equation, the $(m_{ud}, m_s, \mu)$-dependence of the critical temperature is given by
\begin{eqnarray}
\frac{T-T_E}{T_E} = \frac{a_{ud}}{a_t} m_{ud}^{4/5} 
- \frac{a_s}{a_t} \frac{m_E - m_s}{m_E} - \frac{a_{\mu}}{a_t} \mu^2  
\end{eqnarray}
in the critical region, where $a_t, a_s, a_{\mu}$ and $a_{ud}$ are appropriate constants.
Then, at the critical temperature, the critical surface in the $(m_{ud}, m_s, \mu)$ space is described by 
\begin{eqnarray}
c_{ud} m_{ud}^{2/5} + c_s (m_E - m_s) + \mu^2 =0.
\label{eq:mcsur}
\end{eqnarray}
with appropriate constants $c_{ud}$ and $c_s$. 
Hence, the strange quark mass dependence and the $\mu$ dependence of the critical light quark mass $m_{ud}^c$ around the tricritical point are 
\begin{eqnarray}
m_{ud}^c \sim (m_E - m_s)^{5/2}, \hspace{5mm} 
m_{ud}^c \sim \mu^5.
\end{eqnarray}
The first equation describes the critical line on the $\mu=0$ plane sketched in Fig.~\ref{fig:massdep} (right). 
We expect from the first equation that the critical $m_{ud}$ increases very slowly as $m_s$ decreases. 
Similarly, the second equation suggests that the chemical potential dependence of the critical surface $m_{ud}^c (\mu)$ is also small in the low density region, 
since the $\mu$ dependence starts from a term of $\mu^5$ at $m_{ud}=0$. 
This argument is applicable to $(2+ N_{\rm f})$-flavor QCD.

%

\end{document}